\documentclass[referee]{raa}
\usepackage[pagebackref=true]{hyperref}
\usepackage{soul,ulem,color,float,times,lineno,natbib,caption,amssymb,amsmath,siunitx,booktabs,graphicx,newtxtext,threeparttable}
\bibpunct{(}{)}{;}{a}{}{,}

\begin{document}

\title{A Localization Method of High Energy Transients for All-Sky Gamma-Ray Monitor}
\volnopage{ {\bf 20XX} Vol.\ {\bf X} No. {\bf XX}, 000--000}
\setcounter{page}{1}
\author{Yi Zhao$^{*}$\inst{1,2}, Wangchen Xue\inst{2,3}, Shaolin Xiong$^{*}$\inst{2}, Qi Luo\inst{2,3}, Yuanhao Wang\inst{2,3}, Jiacong Liu\inst{2,3}, Heng Yu\inst{1}, Xiaoyun Zhao\inst{2}, Yue Huang\inst{2}, Jinyuan Liao\inst{2}, Jianchao Sun\inst{2}, Xiaobo Li\inst{2}, Qibin Yi\inst{4,2}, Ce Cai\inst{5}, Shuo Xiao\inst{6,7}, Shenglun Xie\inst{8,2}, Chao Zheng\inst{2,3}, Yanqiu Zhang\inst{2,3}, Chenwei Wang\inst{2,3}, Wenjun Tan\inst{2,3}, Zhiwei Guo\inst{9,2}, Chaoyang Li\inst{10,2}, Zhenghua An\inst{2}, Gang Chen\inst{2}, Yanqi Du\inst{11,2}, Min Gao\inst{2}, Ke Gong\inst{2,3}, Dongya Guo\inst{2}, Jiang He\inst{2,3}, Jianjian He\inst{2}, Bing Li\inst{2}, Gang Li\inst{2}, Xinqiao Li\inst{2}, Jing Liang\inst{11,2}, Xiaohua Liang\inst{2}, Yaqing Liu\inst{2}, Xiang Ma\inst{2}, Rui Qiao\inst{2}, Liming Song\inst{2}, Xinying Song\inst{2}, Xilei Sun\inst{2}, Jin Wang\inst{2}, Ping Wang\inst{2}, Xiangyang Wen\inst{2}, Hong Wu\inst{11,2}, Yanbing Xu\inst{2}, Sheng Yang\inst{2}, Dali Zhang\inst{2}, Fan Zhang\inst{2}, Hongmei Zhang\inst{2}, Peng Zhang\inst{12,2}, Shu Zhang\inst{2}, Zhen Zhang\inst{2}, Shijie Zheng\inst{2}, Keke Zhang\inst{13}, Xingbo Han\inst{13}, Haiyan Wu\inst{14}, Tai Hu\inst{14}, Hao Geng\inst{14}, Gaopeng Lu\inst{15}, Wei Xu\inst{16}, Fanchao Lyu\inst{17}, Hongbo Zhang\inst{18}, Fangjun Lu\inst{2}, Shuangnan Zhang\inst{2}}
\institute{
    Department of Astronomy, Beijing Normal University, Beijing 100875, China {\it yi.zhao@mail.bnu.edu.cn} \\
    \and
    Key Laboratory of Particle Astrophysics, Institute of High Energy Physics, Chinese Academy of Sciences, Beijing 100049, China {\it xiongsl@ihep.ac.cn} \\
    \and
    University of Chinese Academy of Sciences, Beijing 100049, China \\
    \and
    School of Physics and Optoelectronics, Xiangtan University, Xiangtan 411105, Hunan, China \\
    \and
    College of Physics, Hebei Normal University, Shijiazhuang 050024, China \\
    \and
    Guizhou Provincial Key Laboratory of Radio Astronomy and Data Processing, Guizhou Normal University, Guiyang 550001, China \\
    \and
    School of Physics and Electronic Science, Guizhou Normal University, Guiyang 550001, China \\
    \and
    Institute of Astrophysics, Central China Normal University, Wuhan 430079, China \\
    \and
    College of physics Sciences Technology, Hebei University, Baoding 071002, China \\
    \and
    Physics and Space Science College, China West Normal University, Nanchong 637002, China \\
    \and
    School of Computing and Artificial Intelligence, Southwest Jiaotong University, Chengdu 611756, China \\
    \and
    College of Electronic and information Engineering, Tongji University, Shanghai 201804, China \\
    \and
    Innovation Academy for Microsatellites of Chinese Academy of Sciences, Shanghai, 201304, China \\
    \and
    National Space Science Center, Chinese Academy of Sciences, Beijing 100190, China \\
    \and
    School of Earth and Space Sciences, University of Science and Technology of China, Hefei 230026, China \\
    \and
    Electronic Information School, Wuhan University, Wuhan 430072, Hubei, China \\
    \and
    Key Laboratory of Transportation Meteorology of China Meteorological Administration, Nanjing Joint Institute for Atmospheric Sciences, Nanjing 210000, China \\
    \and
    Key Laboratory of Middle Atmosphere and Global Environment Observation, Institute of Atmospheric Physics, Chinese Academy of Sciences, Beijing 100029, China \\
    \vs \no
    {\small Received 20XX Month Day; accepted 20XX Month Day}
}

\abstract{
    Fast and reliable localization of high-energy transients is crucial for characterizing the burst properties and guiding the follow-up observations. Localization based on the relative counts of different detectors has been widely used for all-sky gamma-ray monitors. There are two major methods for this counts distribution localization: $\chi^{2}$ minimization method and the Bayesian method. Here we propose a modified Bayesian method that could take advantage of both the accuracy of the Bayesian method and the simplicity of the $\chi^{2}$ method. With comprehensive simulations, we find that our Bayesian method with Poisson likelihood is generally more applicable for various bursts than $\chi^{2}$ method, especially for weak bursts. We further proposed a location-spectrum iteration approach based on the Bayesian inference, which could alleviate the problems caused by the spectral difference between the burst and location templates. Our method is very suitable for scenarios with limited computation resources or time-sensitive applications, such as in-flight localization software, and low-latency localization for rapid follow-up observations.
}
\authorrunning{Zhao et al.}
\titlerunning{High Energy Transients Localization}
\maketitle

\clearpage 
\section{Introduction}

    High-energy transients, e.g. Gamma-ray Bursts (GRBs) \citep{common_GRB_K1973} and Soft Gamma-Ray Repeaters (SGRs) \citep{common_SGR_Woods2004}, are usually first discovered in the gamma-ray band. Fast and reliable localization of these bursts is critically important for joint observation in multi-wavelength and multi-messenger astronomy. For instance, in the case of the first gravitational wave electromagnetic counterpart event (GW 170817) \citep{common_GWGRB_Abbott2017a, common_GWGRB_Abbott2017b, common_GWGRB_Li2018}, the localization of GRB 170817A given by \textit{Fermi} Gamma-ray Burst Monitor (GBM) and INTEGRAL/SPI-ACS \citep{common_GWGRB_Goldstein2017} helped to establish the association between GW 170817 and GRB 170817A, and guide the follow-up observations of this GW source in multi-wavelength.

    Unlike soft X-rays, gamma-rays are very difficult to focus on for imaging. Thus various methods are proposed to localize gamma-ray transients with different kinds of instruments: (1) Counts distribution among detectors used by all-sky Gamma-ray monitors \citep{Loc_KONUS_Mazets1981, Ins_GBM_Meegan2009, Loc_POLAR_Suarez2010, Loc_GBM_Connaughton2015, Loc_GBM_Goldstein2020, Loc_POLAR_Wang2021, Ins_GECAM_Li2022}. (2) Time-delay localization methods (i.e. triangulation) jointly used by multiple spacecraft \citep{Loc_TimeDelay_Hurley2013, Loc_TimeDelay_Xiao2021}. (3) Sky map reconstruction with coded mask imaging \citep{Loc_CodeMask_Preger1999, Loc_CodeMask_Goldwurm2003, Ins_Swift_Krimm2013, Ins_MAXI_Matsuoka2009, Loc_CodeMask_Cordier2015}. (4) Direct measurement of the incident direction of individual photons in relatively high energy (about 10 MeV to GeV) based on Compton scattering \citep{Ins_CGRO2EGRET_Thompson1995} or pair production \citep{Ins_AGILE2GRID_Feroci2007, Ins_LAT_Atwood2009}. (5) Localization with modulation technique used by collimator-based instrument \citep[e.g. \textit{Insight}-HXMT,][]{Loc_Colliamtor_Li2021}.

    Here we focus on the counts distribution localization method for all-sky monitors, i.e. localization based on fitting the source counts in different detectors with different orientations to the template which is the expected counts of burst source from all possible locations. Depending on which statistics to use and how to deal with the burst spectrum, this method could be generally grouped into two approaches: (1) Localization with $\chi^{2}$ minimization and fixed spectral templates, which is represented by the DoL algorithm used by the \textit{Fermi}/GBM \citep{Loc_GBM_Connaughton2015, Loc_GBM_Goldstein2020}. This approach fits the counts distribution in different detectors with several localization templates which are made from several fixed spectra. Obviously, the real spectrum of the burst could usually differ from that of the fixed templates, leading to systematic errors. However, this method is simple and fast, thus widely used in in-flight localization software \citep{Loc_GBM_Connaughton2015, Ins_GECAM_Yun2021}. (2) Localization with the Bayesian inference and fitting the burst location and spectrum simultaneously, represented by the BAyesian Location Reconstruction Of GRBs (BALROG) algorithm \citep{Loc_BALROG_Michael2018} and the MCMC-based localization algorithm developed for Gravitational Wave High-energy Electromagnetic Counterpart All-sky Monitor (GECAM) \citep{Loc_MCMC_Liao2020}. This kind of method is arguably able to give more accurate results than the former, however, it usually requires more computing resources \citep{Loc_BALROG_Michael2018}.

    For the Bayesian-based method, BALROG showed a notable improvement over the DoL algorithm (i.e. $\chi^{2}$ minimization) both in accuracy and precision for localization \citep{Loc_BALROG_Michael2018, Loc_BALROG_Berlato2019}. However, \citet{Loc_GBM_Goldstein2020} compared the accurateness and robustness of the BALROG algorithm and the updated GBM Team's official automated system (RoboBA) based on the DoL algorithm and found that the updated RoboBA is more accurate for the selected GRBs and that there are some technical problems for BALROG algorithm, such as convergence and sensitivity issues.

    There are some scenarios where the computing resource is constrained or time consumption is sensitive, e.g. in-flight localization software and low-latency localization calculation for rapid follow-up observations. The $\chi^{2}$ minimization method, which employs the pixelized sky map and fixed templates, is widely used in these cases since it requires fewer computing resources. However, although $\chi^{2}$ is generally valid when the counts number is large enough (i.e. $>$ 15--20), it will bias when the counts number is few.

    To balance the calculation speed and localization accuracy in the above applications, we propose a modified Bayesian localization method based on the above two types of localization methods, which alleviates the bias from $\chi^{2}$ and the required computing resource is also few as the $\chi^{2}$ method. We also investigate the difference and performance between our Bayesian localization method and the $\chi^{2}$ minimization method with detailed formula derivation and simulations. In order to mitigate the imperfection of the methods with fixed templates, we also propose a location-spectrum iteration approach based on the Bayesian inference.

    This paper is structured as follows: In Section 2, we describe our modified Bayesian localization method, then we revisit the $\chi^{2}$ minimization method. To make a fair comparison, we assume the expectation of the background is precisely known in this paper. These two localization methods are validated in Section 3. In Section 4, The location-spectrum iteration localization strategy is described. Finally, a summary is given in Section 5.

\section{Bayesian Method and $\chi^{2}$ Minimization Method}

    \subsection{Our Bayesian Methods}\label{Sec_Baye}

        In analog to the DoL's $\chi^{2}$ minimization method (see Section \ref{Sec_ChiSq_GBM}) with fixed spectral templates, we modified the Bayesian localization method using the fixed spectral templates with likelihood maximization for each incident direction and with the assumption of the known background \footnote{We note that, in the localization of real observations, the expected value of the background is unknown. We can only obtain the estimated background $\hat{B}$ and its uncertainty $\sigma_{\rm B}$ from background analysis (e.g. the polynomial fitting to the background intervals), and these background uncertainties should be considered in the likelihood of Poisson data. To deal with this case, the Poisson data with Gaussian background (PGSTAT) profile likelihood can be utilized, e.g. \citet{YiZhao_LOC_GECAM}.}. We note that this Bayesian method could invoke different likelihoods. Here we tested both the Poisson likelihood and simplified Gaussian likelihood, as discussed below.

        \subsubsection{Bayesian Method with Poisson Likelihood}\label{Sec_Like_Pois}

            If the background is known precisely, the observational data (i.e. counts in a given detector) follow a simple Poisson distribution:

                \begin{align}
                    P_{\rm Poisson}(S|B,M) = \frac{ (B+M)^S \cdot \exp(-(B+M)) }{ S! }
                    \label{EQ_Pois_R}
                \end{align}
            where $B$ is the expected background counts, $M$ is the expected source counts, and $S$ is the measured counts. The sum of the background $B$ and source $M$ is the expected value of the measured counts. Based on this Poisson distribution, the Poisson likelihood and its logarithmic form for fixed spectral templates localization method can be written as:

                \begin{align}
                    \mathcal L_{\rm P}(i) = \prod \limits_{j=1}^{n} \frac{ \left( b_{j} + f_{i} \cdot m_{j,i} \right) ^ { s_{j} } \cdot \exp( -( b_{j} + f_{i} \cdot m_{j,i} ) ) }{ s_{j} ! }
                    \label{EQ_Pois_P}
                \end{align}

                \begin{align}
                    \ln{ \mathcal L_{\rm P}(i) } = \sum_{j=1}^{n} [ s_{j} \cdot \ln( b_{j} + f_{i} \cdot m_{j,i} ) - ( b_{j} + f_{i} \cdot m_{j,i} ) -\ln s_{j}! ]
                    \label{EQ_Pois_L}
                \end{align}
            where $s_{j}$ is the total observed counts in detector $j$, $n$ is the total number of detectors and $b_{j}$ is the expectation value of the background. Here we use $f_{i} \cdot m_{j,i}$ as the expected source contribution, where $m_{j,i}$ is the localization template of a specific spectrum, which is a matrix of counts of each detector $j$ for each incident direction $i$ (the whole sky is pixelized with HEALPix), and $f_{i}$ is the normalization factor to account for the fluence ratio between the real burst and the preset fixed burst spectrum used to generate the template $m_{j,i}$.

            During the localization process with fixed templates, the $f_{i}$ could be derived from the maximization for each direction ($i$), thus the burst position (i.e. direction $i$) is the only parameter of interested, whose prior could be assumed to be uniform all over the celestial sphere: $P_{\rm prior}(i)=\frac{1}{N}$, where $N$ is the total number of the HEALPix pixels of all sky. In this work, the HEALPix pixels number is set to 41772, i.e. $\sim$ 1$^{\circ}$ for each pixel. With the parameter prior and likelihood as shown in Equation \ref{EQ_Pois_L}, the location results (location center, probability map, and credible region) could be derived through the Bayesian inference.

            We summarize this Bayesian localization method based on Poisson likelihood (denoted as $\mathcal B_{\rm POIS}$ hereafter) as follows:

            \begin{itemize}

                \item[\textbf{Step 1:}] For each incident direction $i$, maximize the likelihood ($ \mathcal L(i) $) by adjusting the normalization factor $f_{i}$. The maximization of likelihood (Equation \ref{EQ_Pois_P}) and logarithmic likelihood (Equation \ref{EQ_Pois_L}) are equivalent for this process.

                \item[\textbf{Step 2:}] Calculate the posterior probability through Bayesian inference. Thus the posterior distribution, $P(i|s)$, could be derived from the prior probability $P_{\rm prior}(i)$, conditional probability for a given direction $i$ to obtain the observed counts $s$ and evidence $P(s)$:

                    \begin{equation}
                    \begin{aligned}
                        P(i|s) &= \frac{ P_{\rm prior}(i) \cdot P(s|i) }{ P(s) }                                     \\
                               &= \frac{ \frac{1}{N} \cdot P(s|i) }{ \sum_{i^{'}} { \frac{1}{N} \cdot P(s|i^{'}) } } \\
                               &= \frac{ P(s|i) }{ \sum_{i^{'}} { P(s|i^{'}) } }
                        \label{EQ_BYS_1}
                    \end{aligned}
                    \end{equation}
                By substituting the conditional probability ($P(s|i)$) with the likelihood ($ \mathcal L(i) $), one can get the posterior probability for each direction ($i$), which is also the localization probability map:

                    \begin{align}
                        P(i) = \frac{ \mathcal L(i) }{ \sum_{i^{'}}{ \mathcal L(i^{'}) } }
                        \label{EQ_BYS_2}
                    \end{align}

                \item[\textbf{Step 3:}] For simplicity, we take the direction with the maximum $P(i)$ as the location center and the Bayesian credible region with $N\%$ highest posterior density (HPD) as the $N\%$ confidence interval of the burst position.

            \end{itemize}

        \subsubsection{Bayesian Method with Gaussian Likelihood}\label{Sec_Like_SG}

            In order to understand the GBM DoL's $\chi^{2}$ method (see Section \ref{Sec_ChiSq_GBM}) in the Bayesian framework, here we structure a Gaussian likelihood. The Gaussian distribution reads:

                \begin{align}
                    P_{\rm Gaussian}(x) = \frac{ 1 }{ \sqrt{ 2 \pi } \cdot \sigma } \cdot \exp( - \frac{ (S-(B+M))^{2} }{ 2 \sigma^{2} } ),
                    \label{EQ_Gaus_R}
                \end{align}
            where $\sigma^{2}$ is the variance. Thus the Gaussian likelihood and its logarithmic form can be written as follows:

                \begin{align}
                    \mathcal L_{\rm G}(i) = \prod \limits_{j=1}^{n} \frac{ 1 }{ \sqrt{2 \pi} \cdot \sigma_{j} } \cdot \exp( - \frac{ (s_{j}-(b_{j}+f_{i} \cdot m_{j,i}))^{2} }{ 2 \sigma_{j}^{2} } )
                    \label{EQ_Gaus_P}
                \end{align}

                \begin{align}
                    \ln{ \mathcal L_{\rm G}(i) } = \sum_{j=1}^{n} [ \ln{ \frac{ 1 }{ \sqrt{2 \pi} \cdot \sigma_{j} } } - \frac{ (s_{j}-(b_{j}+f_{i} \cdot m_{j,i}))^{2} }{ 2 \sigma_{j}^{2} } ]
                    \label{EQ_Gaus_L}
                \end{align}

            Generally, the variance $\sigma_{j}^{2}$ could be either data-dependent or model-dependent. However, to approximate the mathematical form of DoL's $\chi^{2}$ (see Section \ref{Sec_ChiSq_GBM}), here the variance is chosen to be model-dependent:

                \begin{align}
                    \sigma_{j}^{2}=b_{j}+f_{i} \cdot m_{j,i}
                    \label{EQ_Gaus_Var}
                \end{align}
            Parameters are defined the same as in the Poisson case mentioned above.

            Note that the variance is equal to the expectation as the counts follow the Poisson distribution. Using Gaussian distribution to approximate Poisson is generally valid when the number of counts is large (i.e. $>$ 15--20). 

            Because the model-dependent variance term (Equation \ref{EQ_Gaus_Var}) of Equation \ref{EQ_Gaus_P} and \ref{EQ_Gaus_L} for each direction generally approaches $s_{j}$, this term could be dropped out through the maximizing process, Equation \ref{EQ_Gaus_L} thus could be written as a simplified Gaussian likelihood form:

                \begin{align}
                    \ln{ \mathcal L_{\rm SG}(i) } = -\sum_{j=1}^{n} {\frac{ ( s_{j} - (b_{j} + f_{i} \cdot m_{j,i}) )^{2} }{ 2 \sigma_{j}^{2} } }
                    \label{EQ_Gaus_Simp_L}
                \end{align}
            Now, it is clear that maximization of this simplified Gaussian logarithmic likelihood is equivalent to $\chi^{2}$ minimization used by DoL (see Section \ref{Sec_ChiSq_GBM}). From the framework of likelihood, it is explicit that the normalization factor $f_{i}$ used in DoL is an approximate solution that maximizes simplified Gaussian logarithmic likelihood \citep[see Equation \ref{EQ_Gaus_Simp_L} in][]{common_STAT_Blackburn2015}, resembling the approximate solution for minimizing $\chi^{2}$. Owing to the difficulty to obtain the analytical solution of $f_{i}$ in Equation \ref{EQ_Gaus_Simp_L}, we employ the Powell algorithm \citep{common_STAT_Powell1964, common_STAT_Press2007} to numerically calculate the $f_{i}$ for the maximum of the likelihood. This numerical solution can also also be used for the $\chi^{2}$ minimization.

            Once the simplified Gaussian logarithmic likelihood of each incident direction is calculated, the posterior probability and credible region could be derived as Section \ref{Sec_Like_Pois}. This Bayesian method with simplified Gaussian likelihood is denoted as $\mathcal B_{\rm SG}$ hereafter.

    \subsection{ $\chi^{2}$ Minimization Methods }

        \subsubsection{$\chi^{2}$ Minimization with Approximate Solution}\label{Sec_ChiSq_GBM}

            The $\chi^{2}$ employed by GBM team's DoL algorithm is defined as \citep[see Equation A1 and A2 in][]{Loc_GBM_Connaughton2015} :

                \begin{align}
                    \chi^{2}(i) = \sum_{j=1}^{n} { \frac{ ( s_{j} - (b_{j} + f_{i} \cdot m_{j,i}) )^{2} }{ b_{j} + f_{i} \cdot m_{j,i} } }
                    \label{EQ_ChiSq_1}
                \end{align}
            where $s_{j}$ and $b_{j}$ are the total observed and estimated background counts observed in detector $j$ (between 50 and 300 keV for \textit{Fermi}/GBM NaI detectors), respectively, $m_{j,i}$ are the model counts (i.e. localization template) in the same energy range for detector $j$ in direction $i$. The normalization factor $f_{i}$ for direction $i$ is defined as:

                \begin{align}
                    f_{i} = \frac{ \sum_{j=1}^{n} \frac{m_{j,i} \cdot (s_{j}-b_{j})}{s_{j}} }{ \sum_{j=1}^{n} \frac{m_{j,i}^{2}}{s_{j}} }
                    \label{EQ_ChiSq_2}
                \end{align}

            Once the $\chi^{2}$ for the whole sky map is calculated, the contour of $\Delta \chi^{2}=C$ is regarded as the $N\%$ statistical error region, where $C$ is the Percent Point Function (PPF) of $\chi^{2}$ distribution with degree of freedom 2 for $N\%$, i.e. $\Delta \chi^{2}=2.3$ represented the $68\%$ statistical uncertainty.

            From the comparison between this $\chi^{2}$ method and the above Bayesian method ($\mathcal B_{\rm SG}$), the normalization factor $f_{i}$ used by DoL is just an approximate solution to minimize $\chi^{2}$ (see Section \ref{Sec_Like_SG}). Also, this $\chi^{2}$ does not consider the uncertainties of the estimated background. Furthermore, the large number of counts is implicitly assumed since Gaussian distribution is used.

            This $\chi^{2}$ method used by the GBM DoL algorithm is denoted as $\chi_{\rm GBM}^{2}$ hereafter.

        \subsubsection{$\chi^{2}$ Minimization with Numerical Solution}\label{Sec_ChiSq_MIN}

            As mentioned above, Equation \ref{EQ_ChiSq_2} is an approximate solution to minimize $\chi^{2}$ and it could be accurately calculated by numerical solution. Thus we studied a $\chi^{2}$ statistic in which the normalization factor $f_{i}$ comes from a numerical solution and other calculations are the same as Section \ref{Sec_ChiSq_GBM}. This $\chi^{2}$ method is denoted as $\chi_{\rm MIN}^{2}$ in this paper.

            The main technical details of these four localization methods using different statistical frameworks mentioned in this chapter are summarized in Table \ref{TABLE_Scheme}.

    %

    \begin{table*}[ht]
        \centering
        \caption{ The localization methods described in Section 2. }
        \label{TABLE_Scheme}
        \begin{threeparttable}
            \begin{tabular}{cccccc}
                Abbreviation & Localization Method & Statistics & Error Region Estimation\tnote{1} & Description \\
                \hline
                $\mathcal B_{\rm POIS}$ & Bayesian with Poisson Likelihood & Poisson Likelihood & HPD\tnote{2} Credible Region & Section \ref{Sec_Like_Pois} \\
                $\mathcal B_{\rm SG}$ & Bayesian with Gaussian Likelihood & Gaussian Likelihood & HPD Credible Region & Section \ref{Sec_Like_SG} \\
                $\chi_{\rm GBM}^{2}$ & $\chi^{2}$ Minimization with Approximate Solution & $\chi^{2}$ & $\Delta \chi^{2}$ & Section \ref{Sec_ChiSq_GBM} \\
                $\chi_{\rm MIN}^{2}$ & $\chi^{2}$ Minimization with Numerical Solution & $\chi^{2}$ & $\Delta \chi^{2}$ & Section \ref{Sec_ChiSq_MIN} \\
                \hline
            \end{tabular}
            \begin{tablenotes}
                \footnotesize
                \item[1] The error estimation is for the localization error region.
                \item[2] Highest Posterior Density (HPD).
            \end{tablenotes}
        \end{threeparttable}
    \end{table*}

\clearpage
\section{Comparison and Validation}

    To quantitatively evaluate and compare the above four localization methods, we conduct a Monte Carlo (MC) simulation\footnote{We note that the real observation suffers from unknown systematic errors and is thus not suitable for this test.} to make tests.

    It should be noticed that several treatments are employed for simplicity and clarity: (1) The expectation value of the background is assumed to be known which allows us to eliminate the influence of background uncertainties. Such an effect is trivial for the present comparison study. (2) The \textit{Fermi}/GBM detector configuration (i.e. 12 NaI detectors) and instrumental response are adopted, however, these localization methods are applicable for any other all-sky monitors of a similar design, such as GECAM.

    Key parameters (such as the position and spectrum of the burst source) used in the simulation are listed in Table \ref{TABLE_SourceType}. The simulated counts in each detector are derived from the Poisson fluctuation of the total expected counts, which are the expected counts of source contribution (i.e. burst spectrum convolved with the detector response) plus the expected background.

    \begin{table*}[ht]
        \centering
        \caption{ Characteristics of the burst used in the localization simulation. The incident angle is $\rm Zenith=5.85^{\circ}$, $\rm Azimuth=22.50^{\circ}$ in GBM's spacecraft coordinates which corresponds to $\rm RA=184.65^{\circ}$, $\rm DEC=-67.72^{\circ}$ (true position) at 2021-01-01T01:00:00 UTC. The fixed background level is set to 1000 counts/s for each detector. Fluence is calculated in 10-1000 keV. }
        \label{TABLE_SourceType}
        \begin{tabular}{cccccc}
            Source Intensity Type          & Medium Bright Burst \\
            \hline
            Spectral Model                 & Comptonized         \\
            Spectral Index                 & -1.50               \\
            $E_{\rm peak}$ (keV)           & 200                 \\
            Duration (s)                   & 10.0                \\
            Fluence $(\rm erg/\rm cm^{2})$ & \num{1.2e-5}        \\
            \hline
        \end{tabular}
    \end{table*}

\begin{figure*}[ht]
        \flushleft

        \begin{minipage}{0.24\linewidth}
            \centering
            \subfloat[][]{\includegraphics[width=4.0cm]{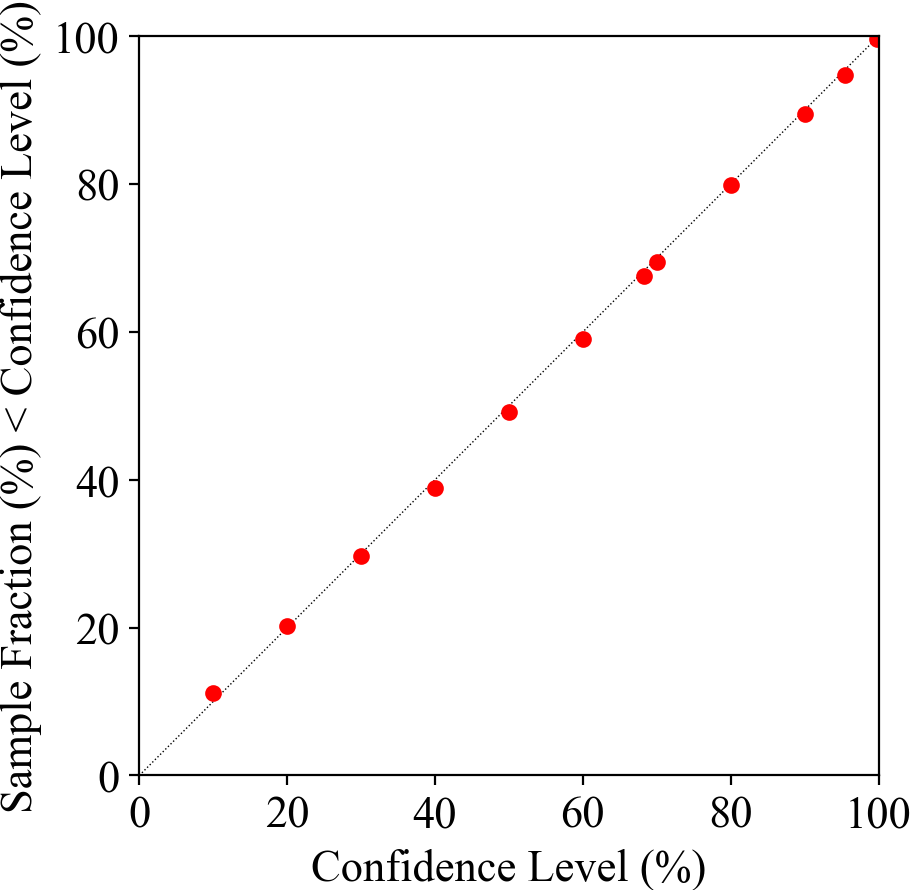}}
            \subfloat[][]{\includegraphics[width=4.0cm]{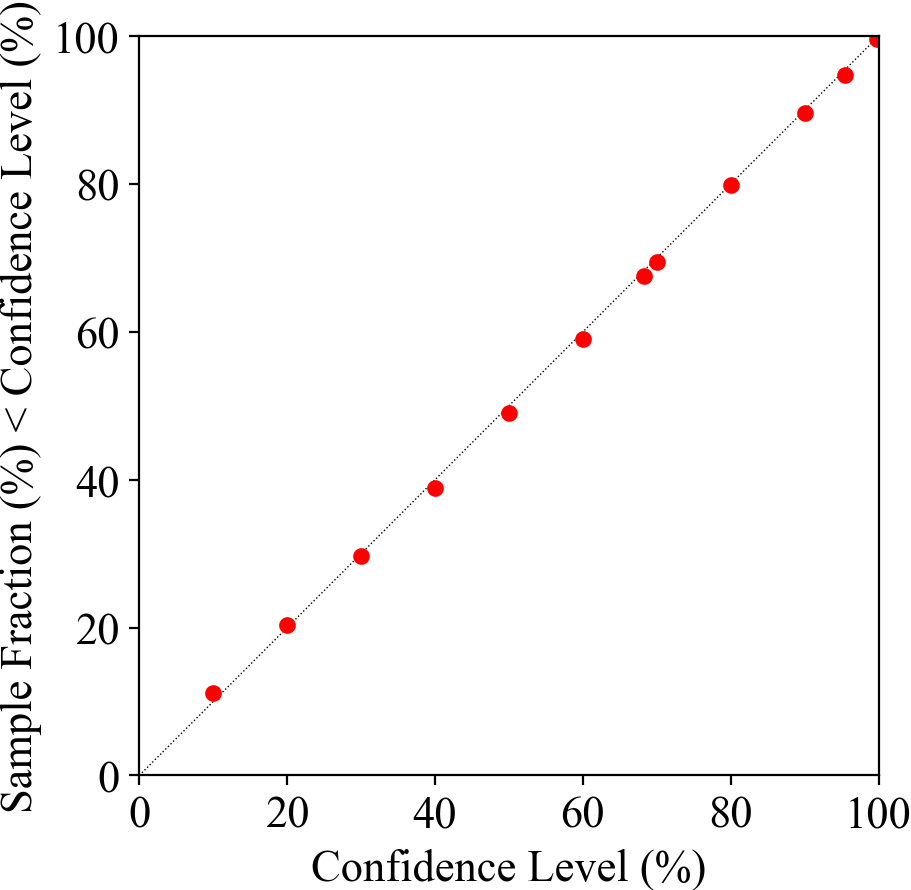}}
            \subfloat[][]{\includegraphics[width=4.0cm]{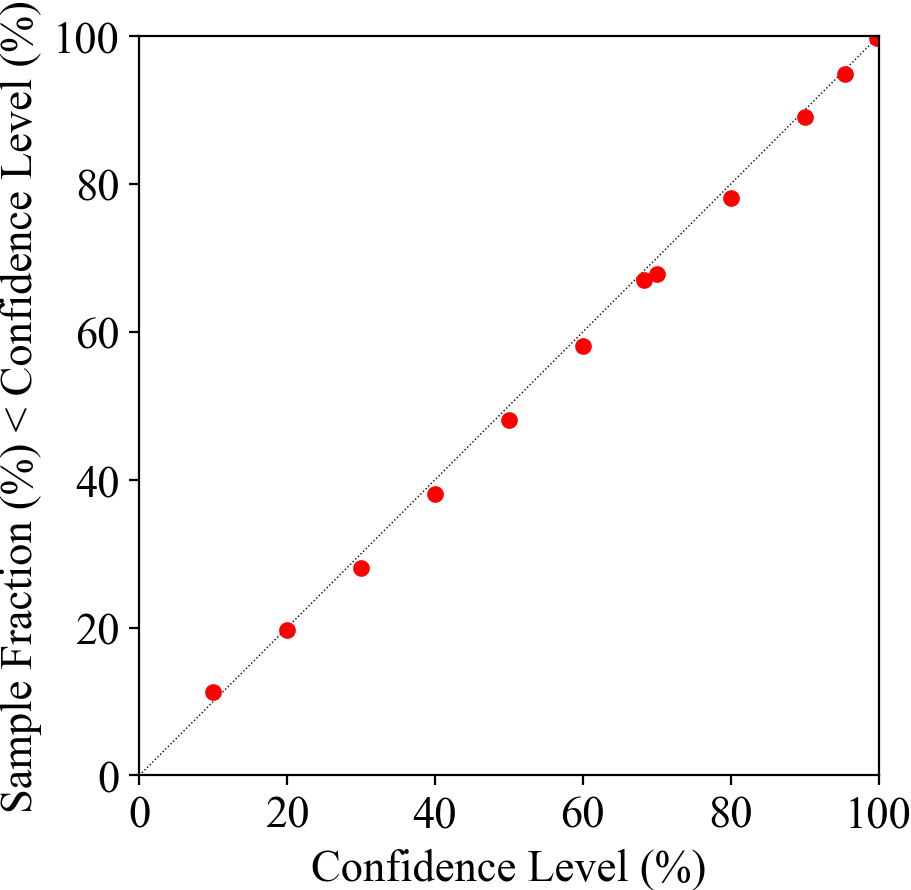}}
            \subfloat[][]{\includegraphics[width=4.0cm]{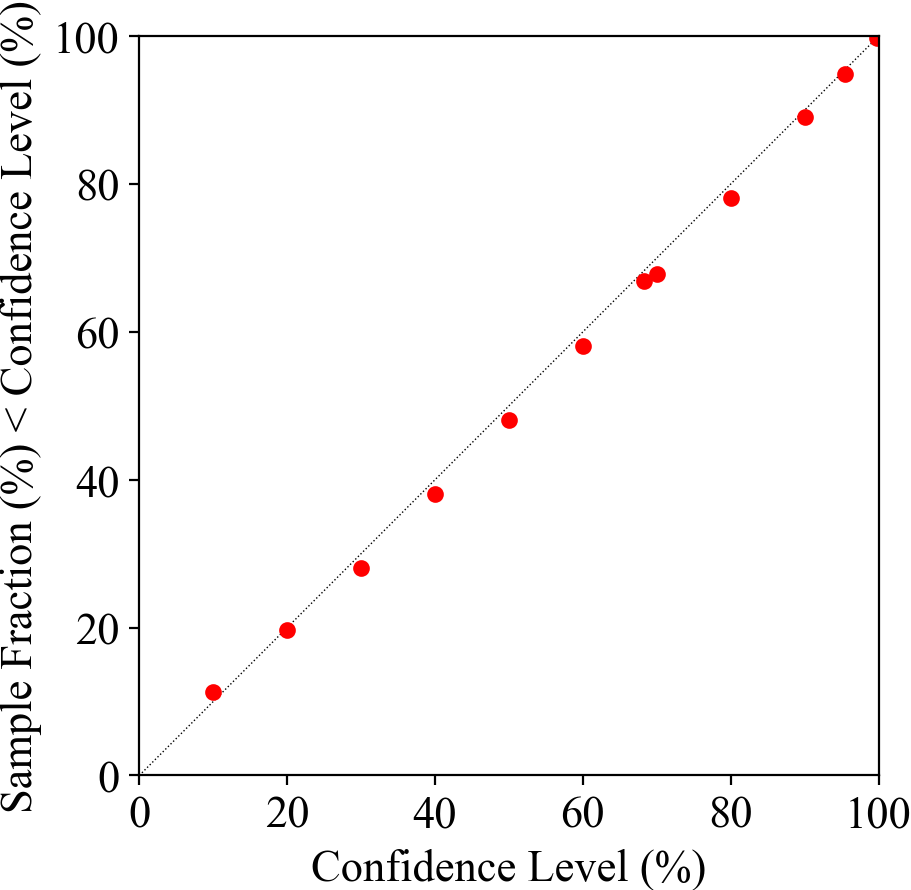}}
        \end{minipage}

        \begin{minipage}{0.24\linewidth}
            \centering
            \subfloat[][]{\includegraphics[width=4.0cm]{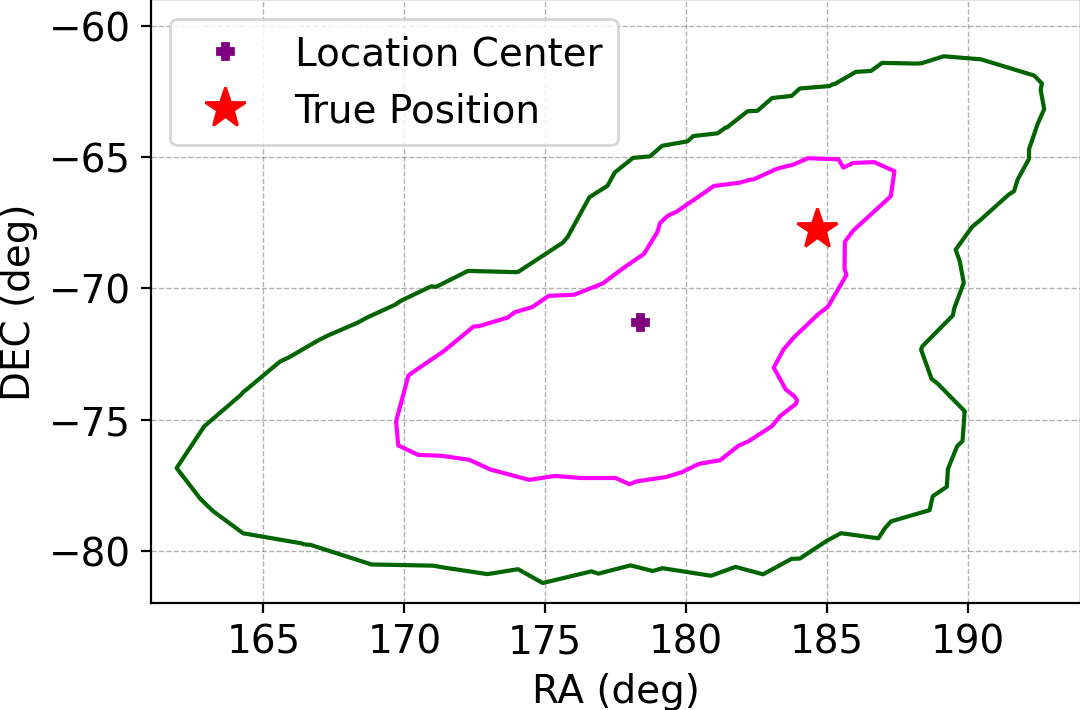}}
            \subfloat[][]{\includegraphics[width=4.0cm]{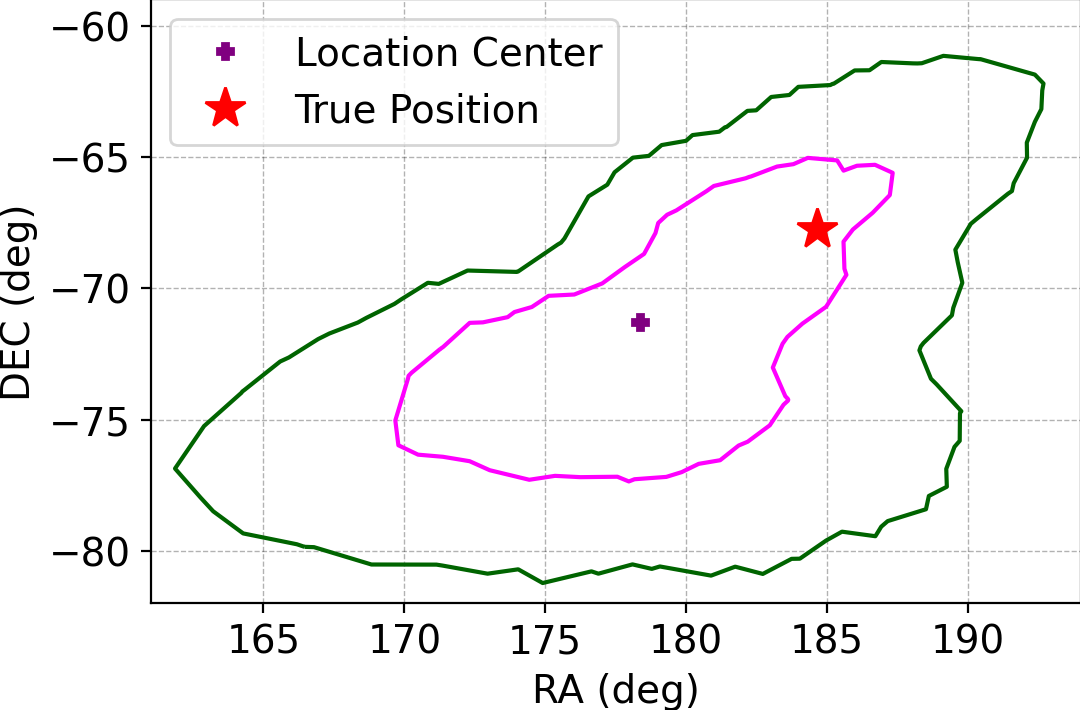}}
            \subfloat[][]{\includegraphics[width=4.0cm]{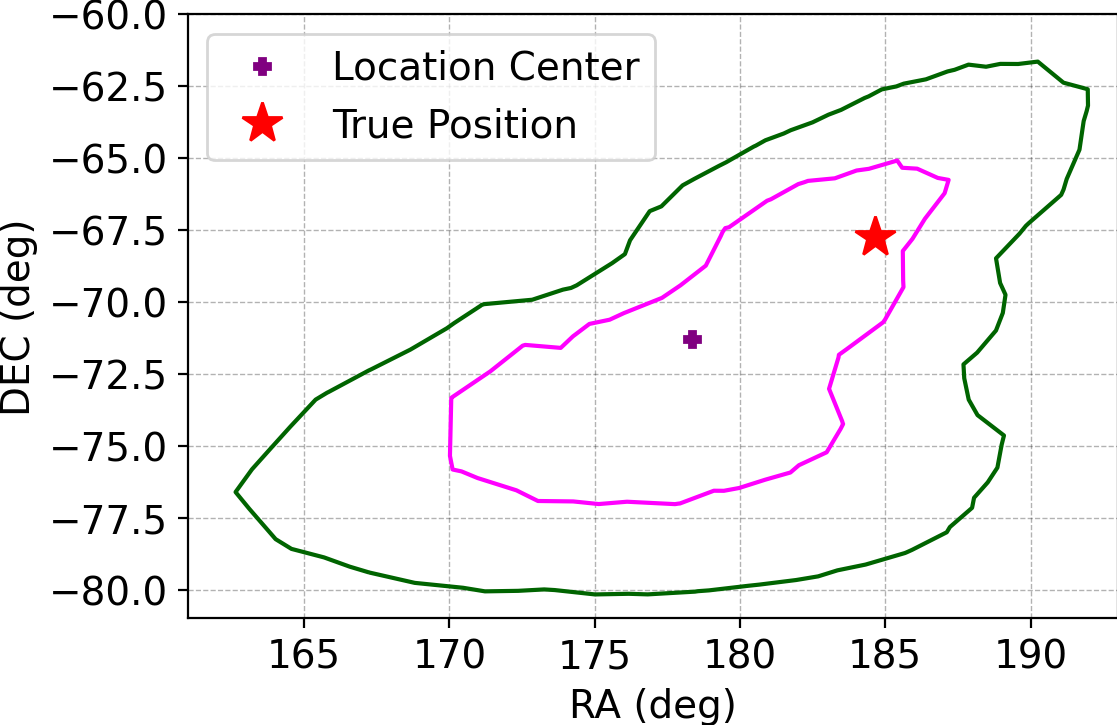}}
            \subfloat[][]{\includegraphics[width=4.0cm]{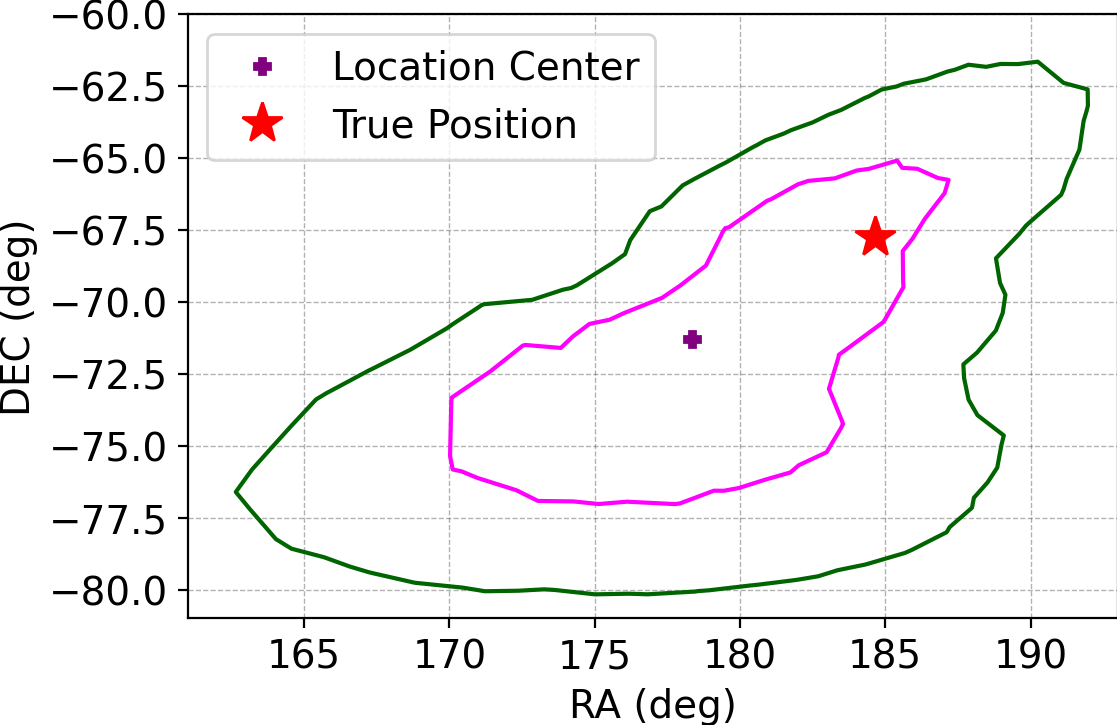}}
        \end{minipage}

        \caption{ Test results for medium bright bursts (Table \ref{TABLE_SourceType}) with simulations. The upper panels (a to d) show the statistical results (i.e. the fraction of bursts with the cumulative probability $N\%$ of true position \textless the corresponding confidence level $N\%$) for four localization methods (see Table \ref{TABLE_Scheme}): (a) $\mathcal B_{\rm POIS}$ (see Section \ref{Sec_Like_Pois}), (b) $\mathcal B_{\rm SG}$ (see Section \ref{Sec_Like_SG}), (c) $\chi_{\rm MIN}^{2}$ (see Section \ref{Sec_ChiSq_MIN}), (d) $\chi_{\rm GBM}^{2}$ (see Section \ref{Sec_ChiSq_GBM}). The dashed line represents the one-one line. The confidence level is 10\% to 90\% step by 10\% as well as 68.27\%, 95.45\%, and 99.73\%. The lower panels (e to h) show inspections of individual simulated bursts for the same localization method as the corresponding upper panels. The magenta and green lines mark 68.27\% and 95.45\% HPD credible regions, respectively. The purple cross and red star represent the location center and true position, respectively. }
        \label{fig1}
    \end{figure*}

    The localization simulation results for the medium bright burst (see Table \ref{TABLE_SourceType} for burst parameters) are shown in Figure \ref{fig1}. To validate the location probability map and credible region, we check the distribution of the real burst position's cumulative probability in the location maps for simulated bursts. This distribution check and the detailed inspections of location maps for individual simulated bursts show that all these four localization methods based on Bayesian and $\chi^{2}$ minimization can give consistent and correct location results (especially the localization error region) for medium bright bursts. This finding is understandable because the medium bright bursts could give a large number of counts in detectors, the Gaussian distribution could well approximate the Poisson distribution, and the $\Delta \chi^{2}$ could be used to derive the confidence region.

    However, we find that some localization methods would fail to give a correct probability map when the burst becomes weak to some extent. Here we explore the burst intensity threshold where these methods give generally reliable localization results for two different cases: background-dominant case and source-dominant case.

\begin{figure*}[ht]
        \flushleft

        \begin{minipage}{0.24\linewidth}
            \centering
            \subfloat[][]{\includegraphics[width=4.0cm]{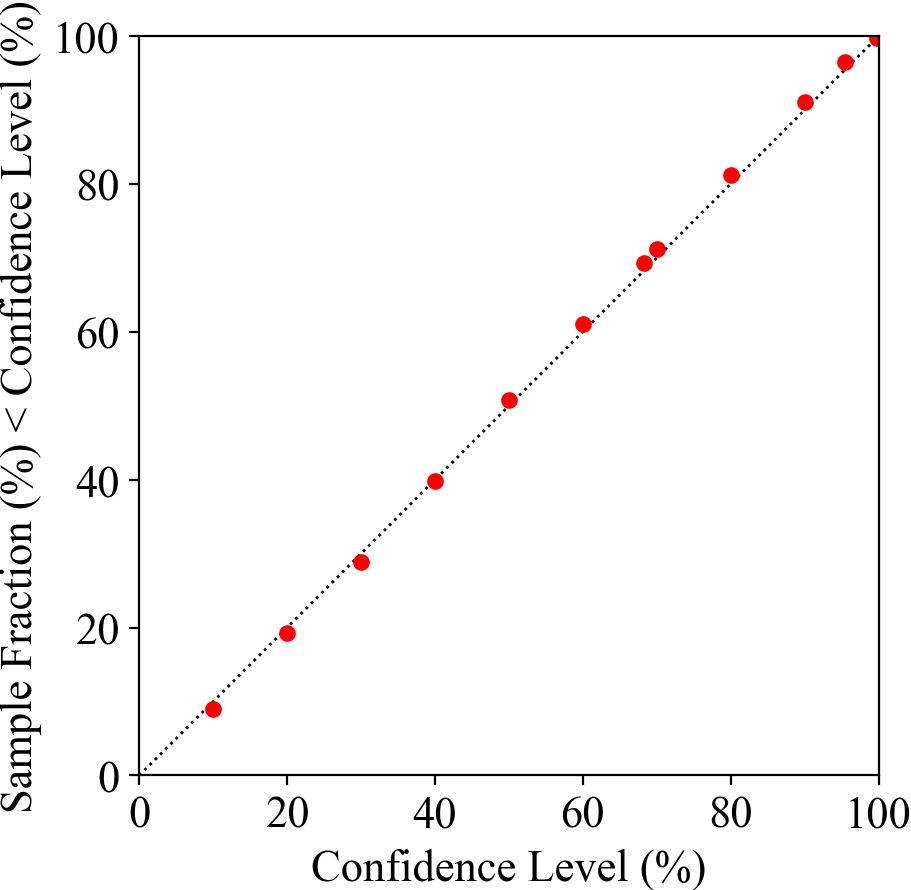}}
            \subfloat[][]{\includegraphics[width=4.0cm]{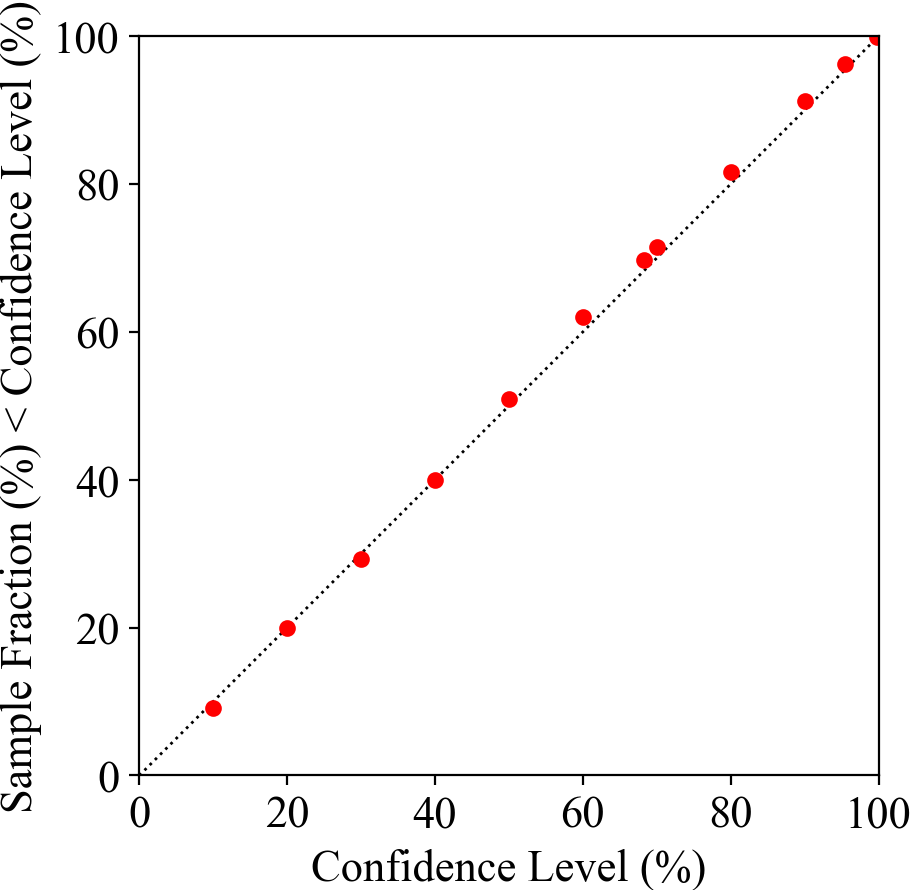}}
            \subfloat[][]{\includegraphics[width=4.0cm]{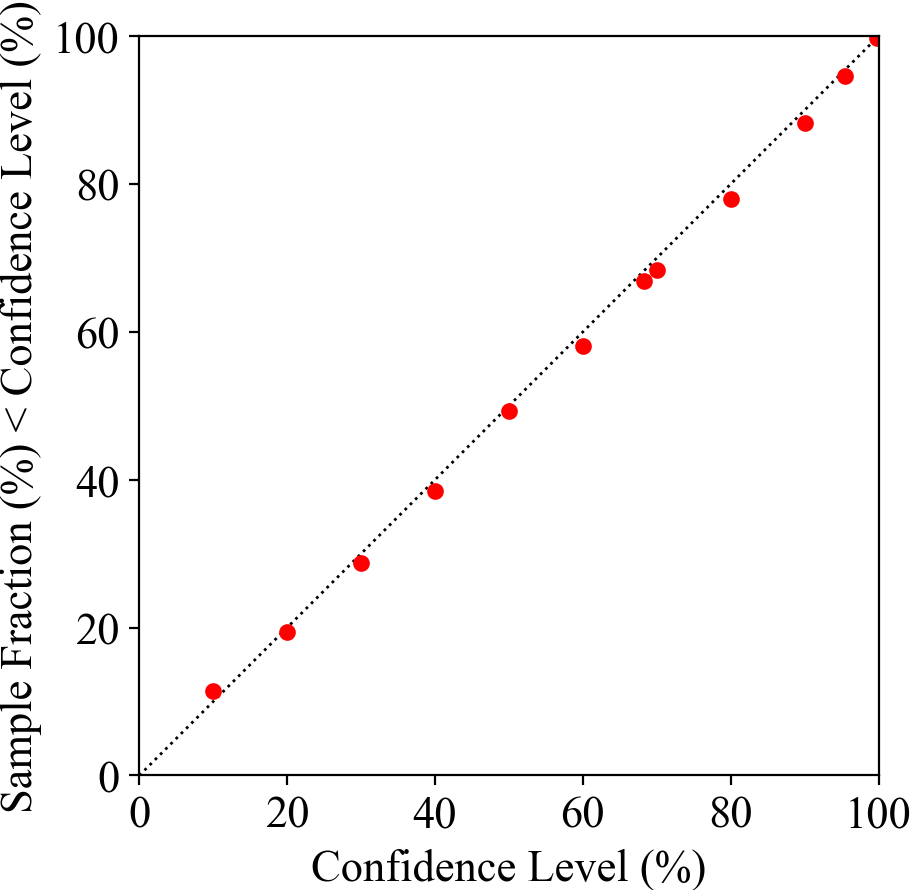}}
            \subfloat[][]{\includegraphics[width=4.0cm]{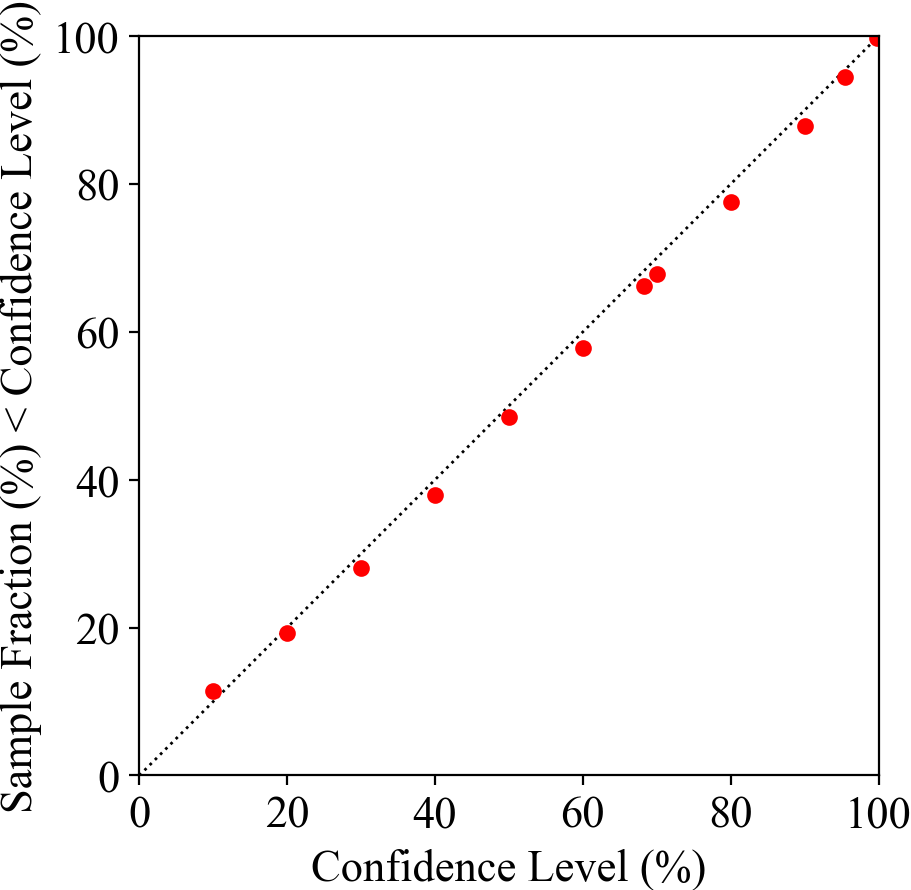}}
        \end{minipage}

        \begin{minipage}{0.24\linewidth}
            \centering
            \subfloat[][]{\includegraphics[width=4.0cm]{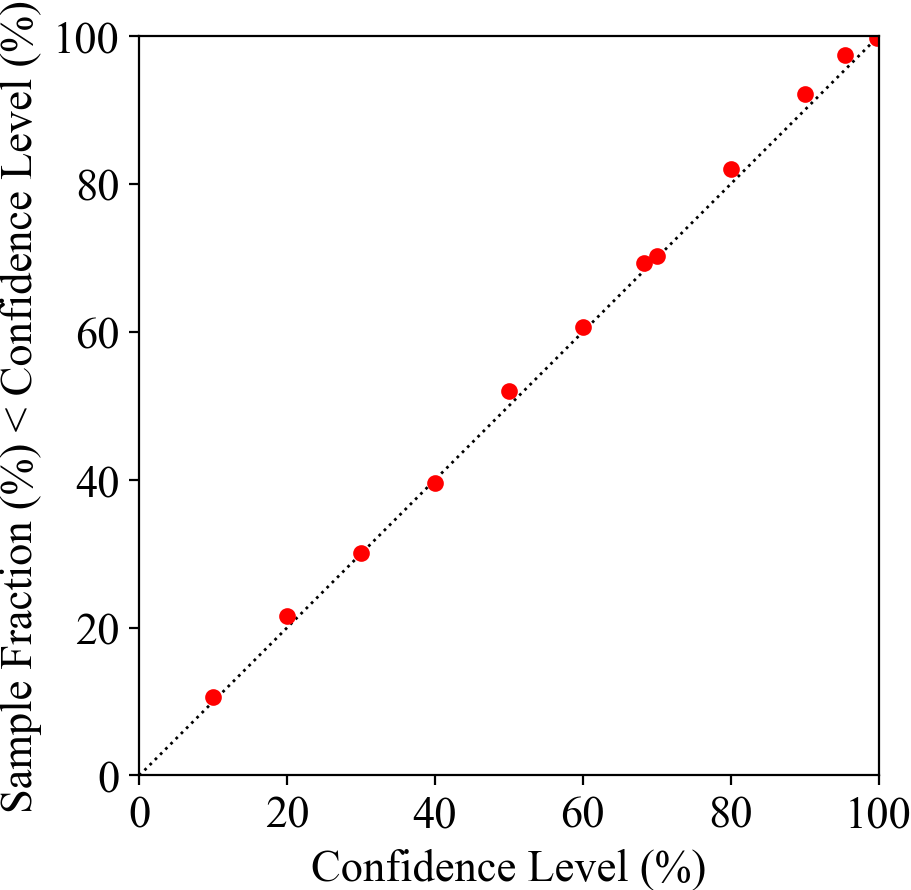}}
            \subfloat[][]{\includegraphics[width=4.0cm]{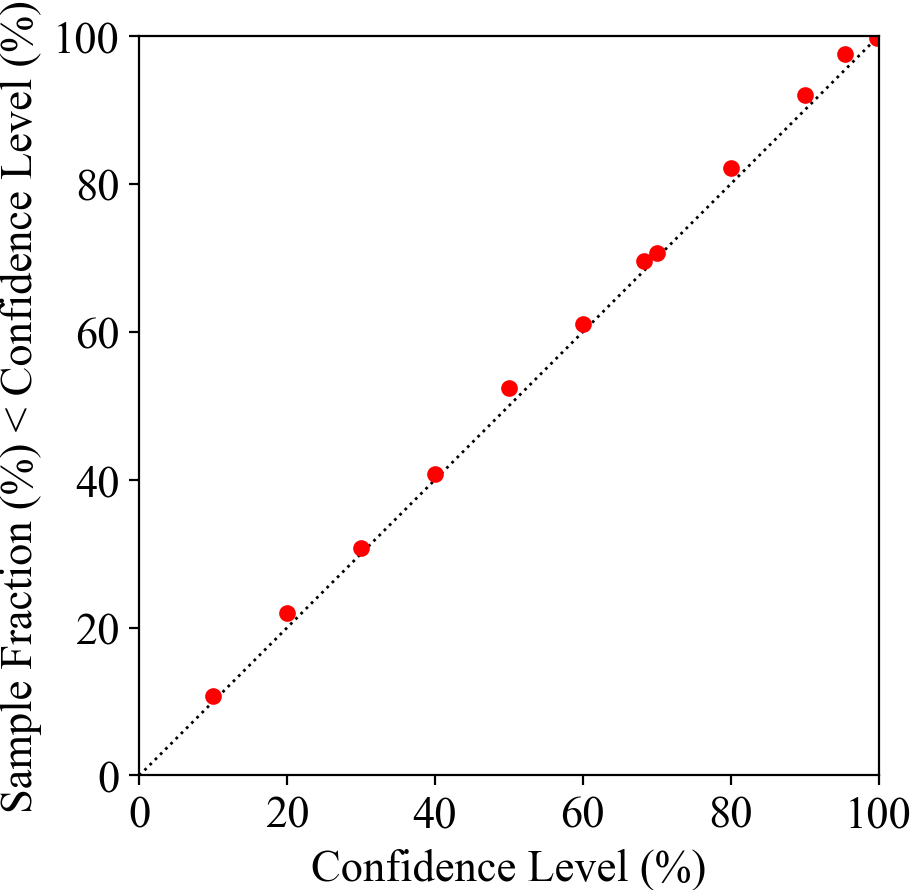}}
            \subfloat[][]{\includegraphics[width=4.0cm]{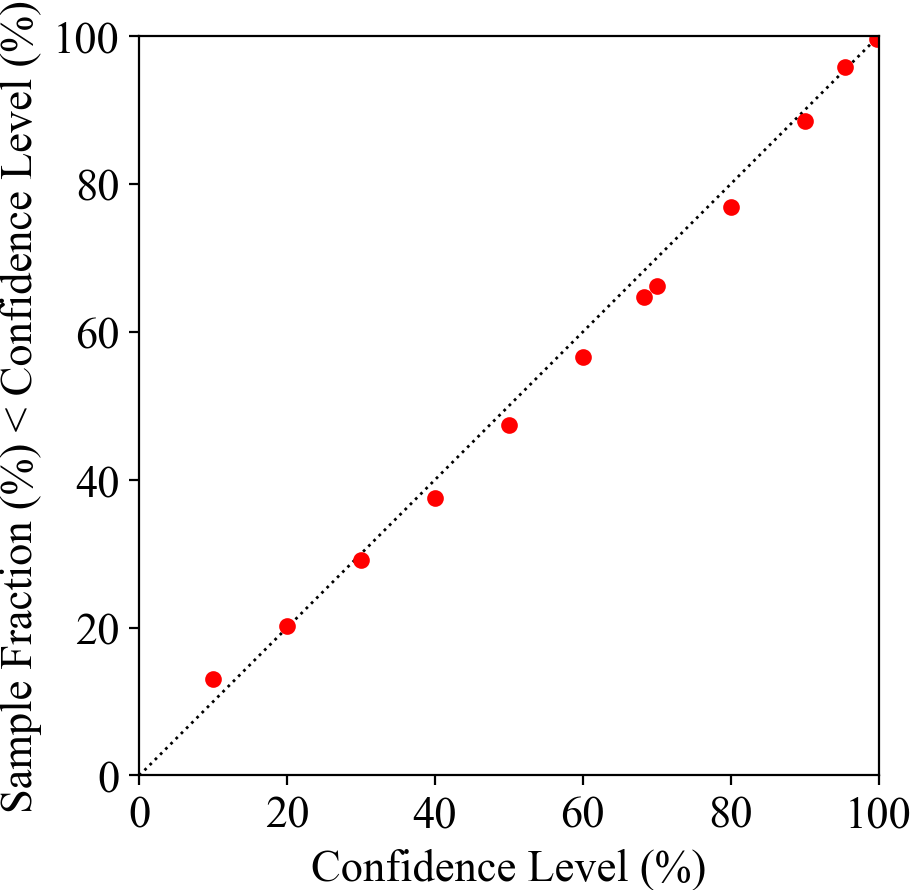}}
            \subfloat[][]{\includegraphics[width=4.0cm]{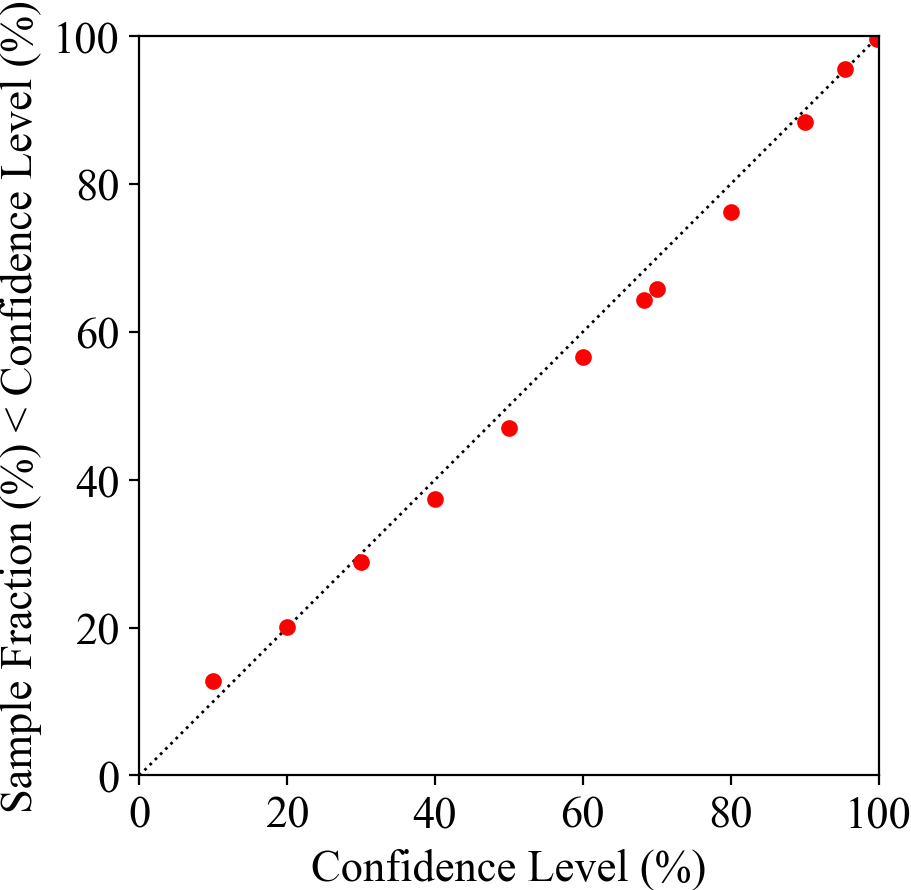}}
        \end{minipage}

        \begin{minipage}{0.24\linewidth}
            \centering
            \subfloat[][]{\includegraphics[width=4.0cm]{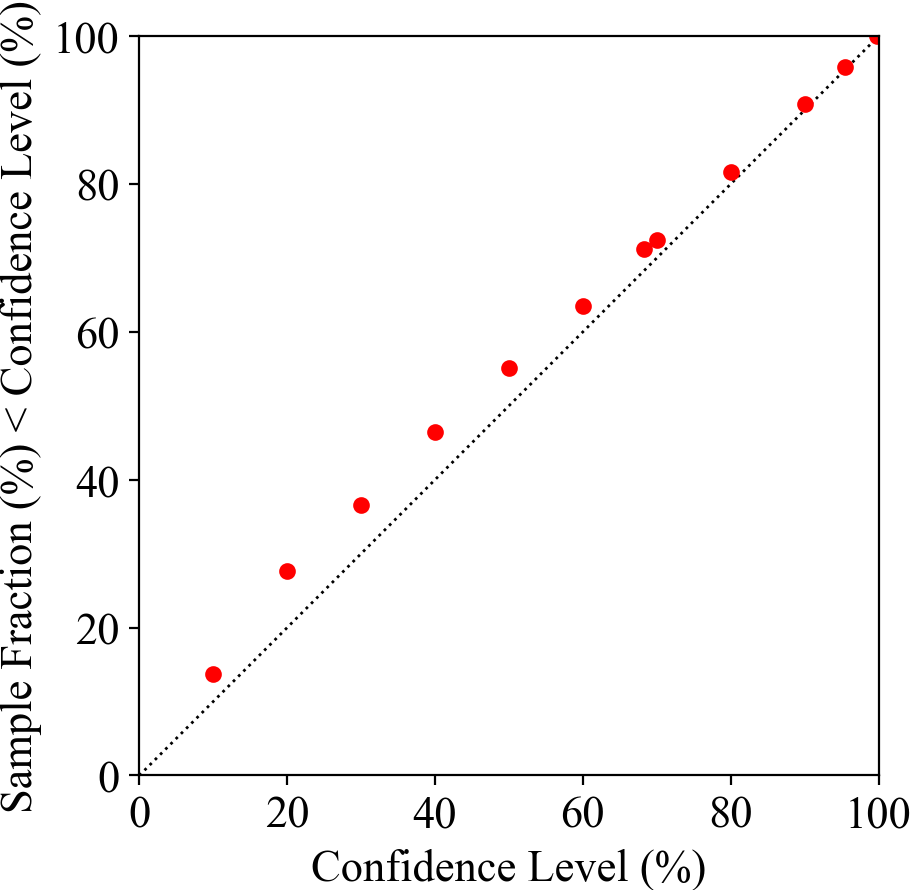}}
            \subfloat[][]{\includegraphics[width=4.0cm]{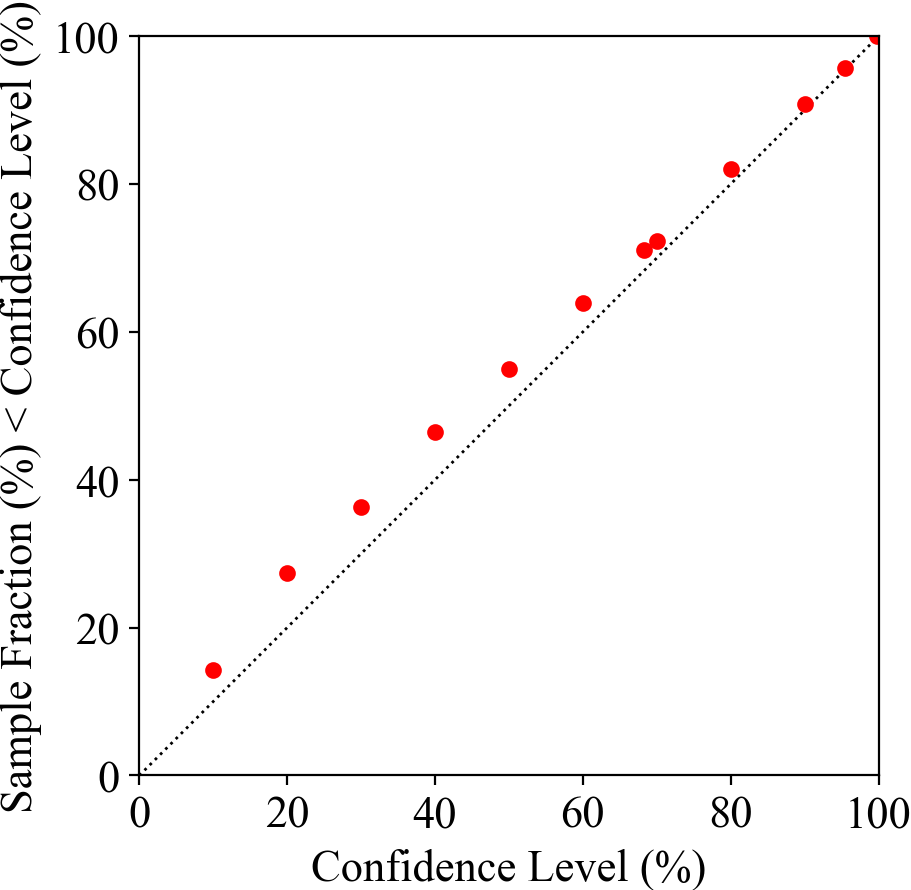}}
            \subfloat[][]{\includegraphics[width=4.0cm]{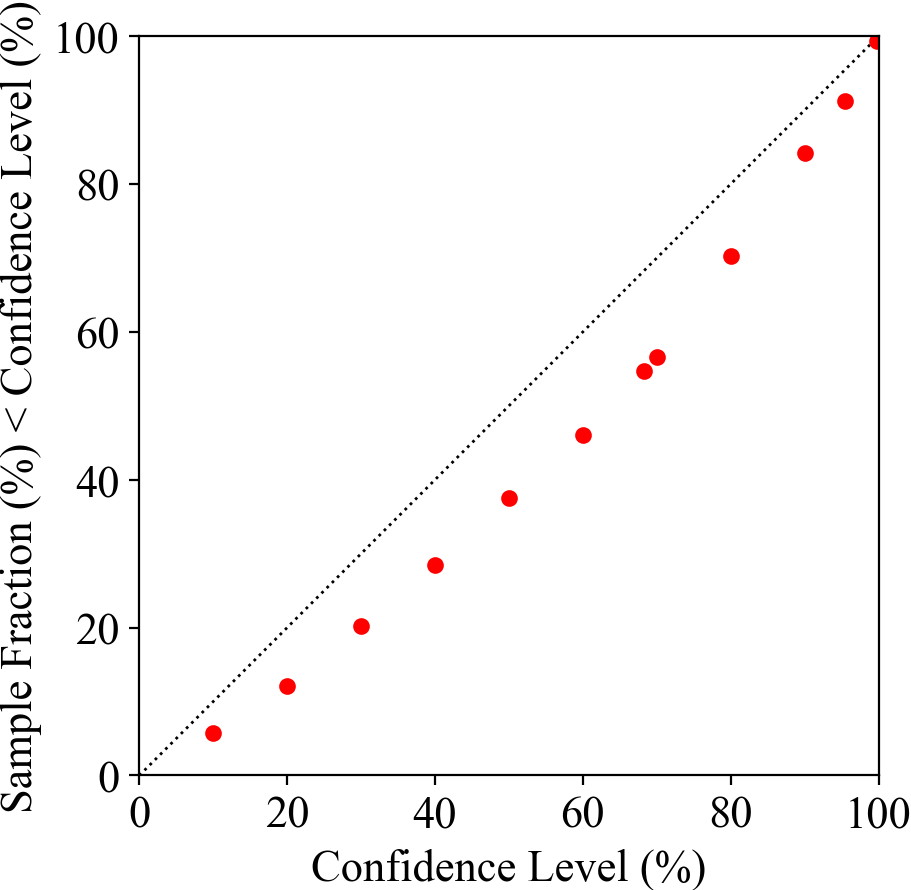}}
            \subfloat[][]{\includegraphics[width=4.0cm]{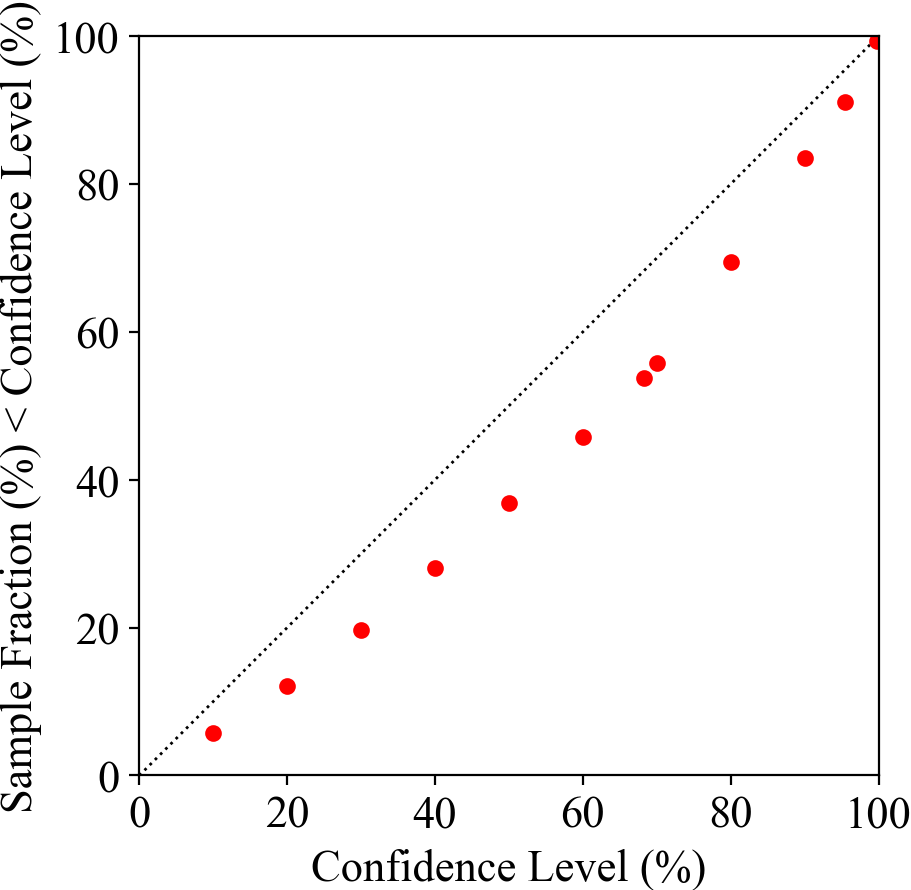}}
        \end{minipage}

        \caption{ Test results for background-dominating weak bursts. The upper panels show the localization results obtained with different localization methods at Signal-to-Noise Ratio (SNR)=9, including (a) $\mathcal B_{\rm POIS}$ (see Section \ref{Sec_Like_Pois}), (b) $\mathcal B_{\rm SG}$ (see Section \ref{Sec_Like_SG}), (c) $\chi_{\rm MIN}^{2}$ (see Section \ref{Sec_ChiSq_MIN}), (d) $\chi_{\rm GBM}^{2}$ (see Section \ref{Sec_ChiSq_GBM}). Here SNR is defined as $\frac{\rm Net Counts}{\sqrt{\rm Background}}$ for the brightest 3 detectors. The medium panels show the corresponding localization results obtained with different methods mentioned in the upper panels at SNR=8 and the lower panels show those results at SNR=3. Other captions are the same as Figure \ref{fig1}. The localization statistical error is reasonable for all methods at SNR=9. $\chi_{\rm MIN}^{2}$ and $\chi_{\rm GBM}^{2}$ start to deviate one-one line at SNR=8 and then more severe at SNR=3. For SNR=3, two Bayesian methods overestimate localization error regions while two $\chi^{2}$ minimization localization methods underestimate localization error regions. Although all localization methods cannot give a correct probability map \textless 90\% confidence level, $\mathcal B_{\rm POIS}$ and $\mathcal B_{\rm SG}$ still could give a correct localization probability map at \textgreater 90\% HPD credible region. }
        \label{fig4}
    \end{figure*}

    For the background-dominant case (i.e. the background counts are much more than burst source counts), we take the GRB 170817A as the input burst. The time-integrated spectrum of GRB 170817A \citep{common_GWGRB_Goldstein2017} is adopted (the total source counts in the brightest 3 detectors $\sim 900$) and the background level is set to 600 counts/detector in our simulations. We verified the localization results as the input burst intensity decreases. Here we define the Signal-to-Noise Ratio (SNR) as $\frac{\rm Net Counts}{\sqrt{\rm Background}}$ for the 3 brightest detectors. As shown in Figure \ref{fig4}, when the SNR decreases to 9, all localization results of $\mathcal B$ and $\chi^{2}$ are basically around the one-one line. But when SNR is $\sim$ 8 (i.e. the summed source counts in the brightest 3 detectors $\sim 430$ and the input burst intensity decreases to $\sim 38\%$ of GRB 170817A), the two $\chi^{2}$ methods start to deviate one-one line, while the two Bayesian methods still could give a correct localization probability map.

    When the burst intensity decreased to SNR=3 (the summed source counts in the brightest 3 detectors $\sim 130$ which means $\sim$14\% of GRB 170817A), all methods failed to give a correct localization probability map. These two Bayesian methods overestimate localization error regions while two $\chi^{2}$ methods underestimate error regions. And the Bayesian method seems closer to the one-one line than the two $\chi^{2}$ methods. Although it cannot give a correct probability map \textless 90\% confidence level (CL), we note that $\mathcal B_{\rm POIS}$ and $\mathcal B_{\rm SG}$ could give a correct credible region at \textgreater 90\% CL, i.e. 95.45\% and 99.73\% CL. As the detection horizon of gravitational detectors increases, the detected GW events would be further and the GW-associated GRBs might be weaker, say $\sim 14\%$ to $\sim 30\%$ of GRB 170817A, for which the $\chi^{2}$ methods will underestimate the location error, thus we suggest that a credible region \textgreater 90\% CL of $\mathcal B_{\rm POIS}$ and $\mathcal B_{\rm SG}$ should be used to estimate the location error of these weak bursts.

\begin{figure*}[ht]
        \flushleft

        \begin{minipage}{0.24\linewidth}
            \centering
            \subfloat[][]{\includegraphics[width=4.0cm]{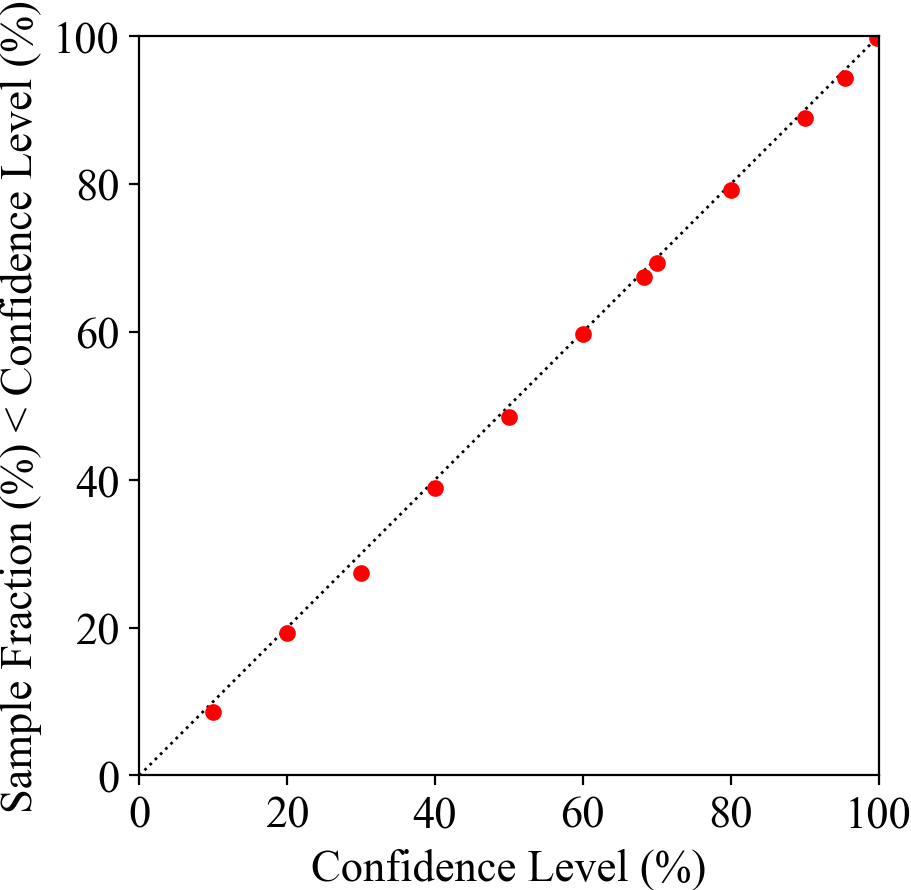}}
            \subfloat[][]{\includegraphics[width=4.0cm]{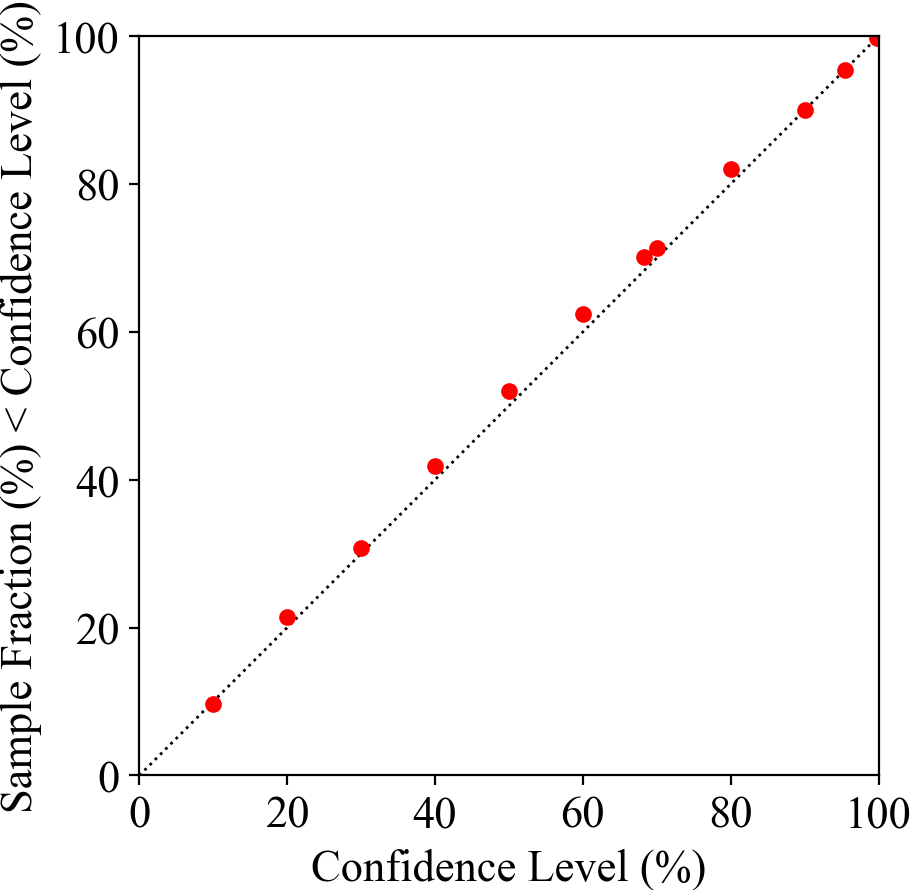}}
            \subfloat[][]{\includegraphics[width=4.0cm]{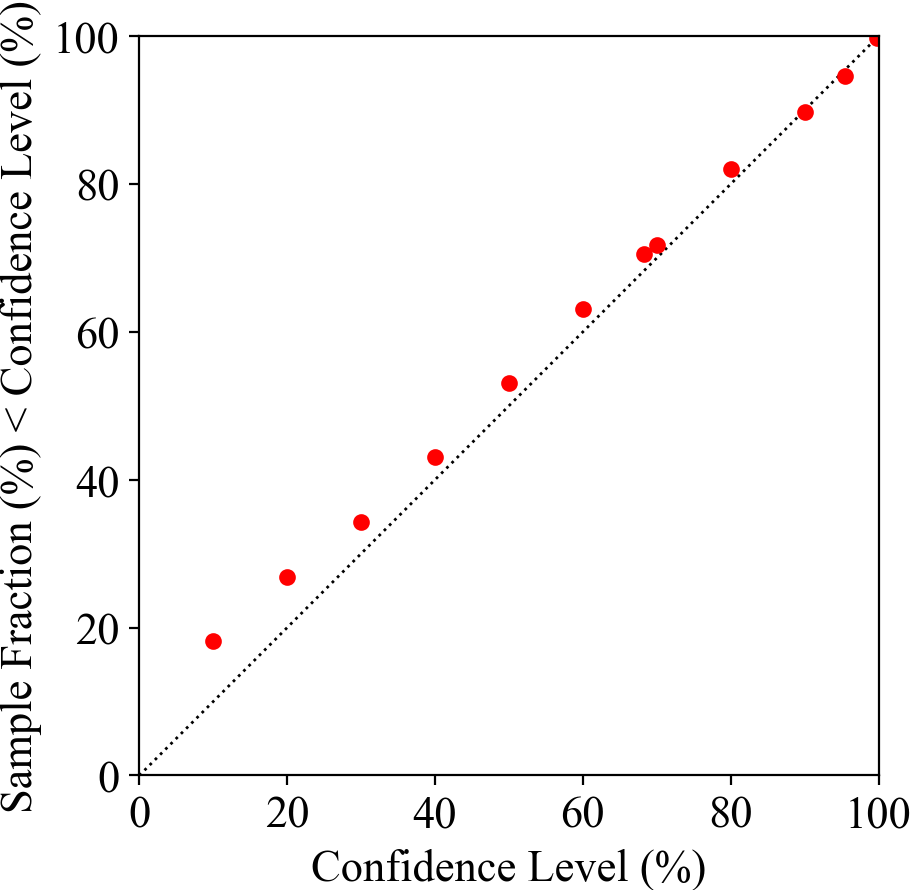}}
            \subfloat[][]{\includegraphics[width=4.0cm]{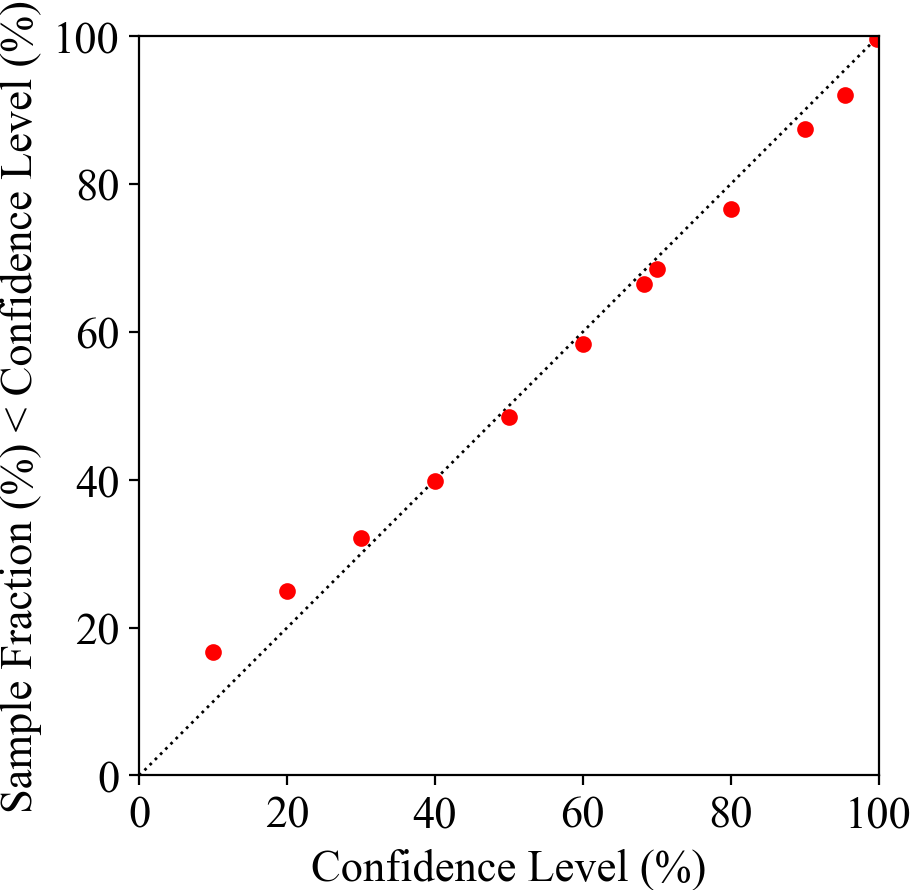}}
        \end{minipage}

        \begin{minipage}{0.24\linewidth}
            \centering
            \subfloat[][]{\includegraphics[width=4.0cm]{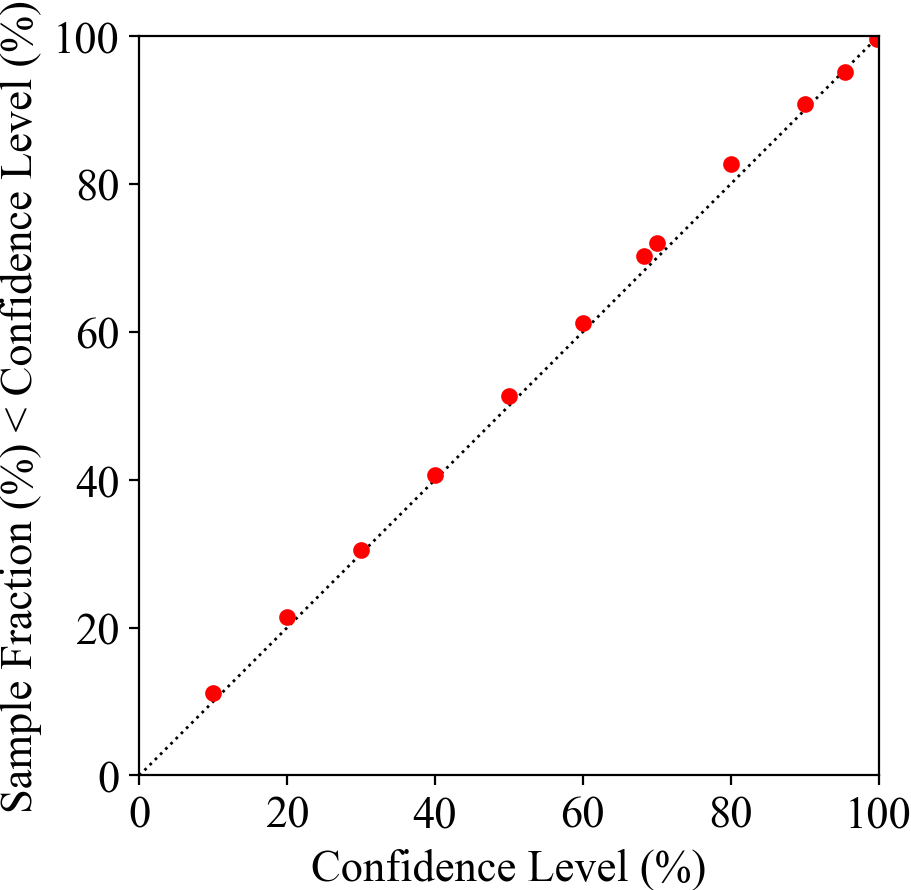}}
            \subfloat[][]{\includegraphics[width=4.0cm]{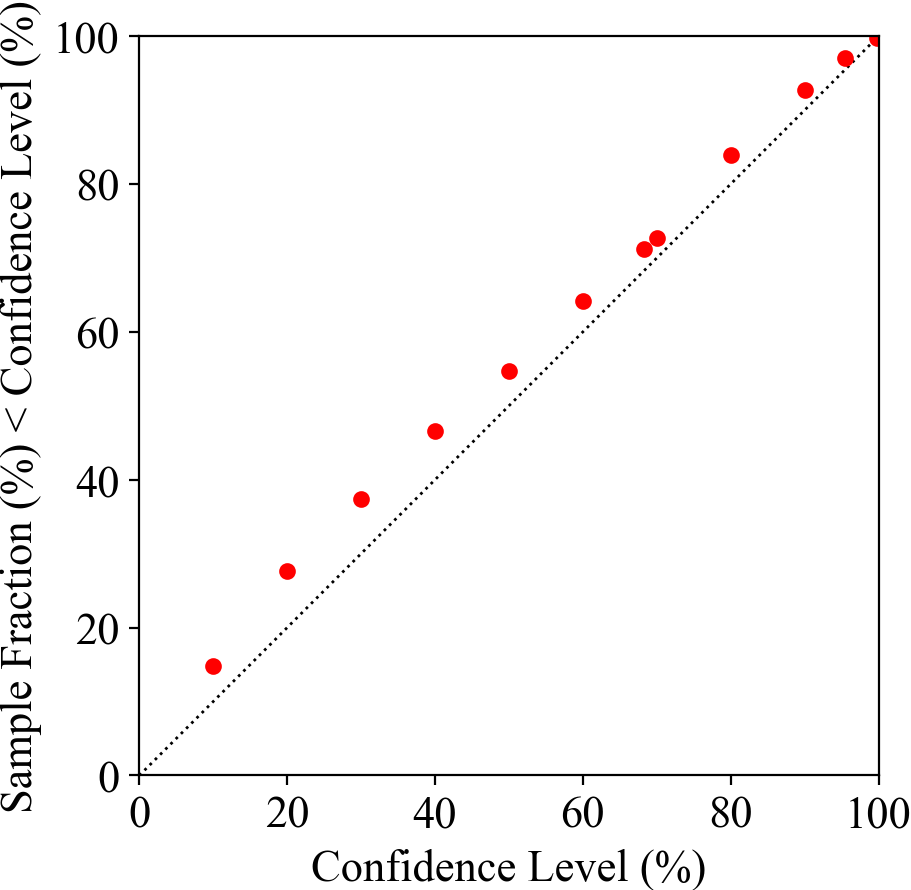}}
            \subfloat[][]{\includegraphics[width=4.0cm]{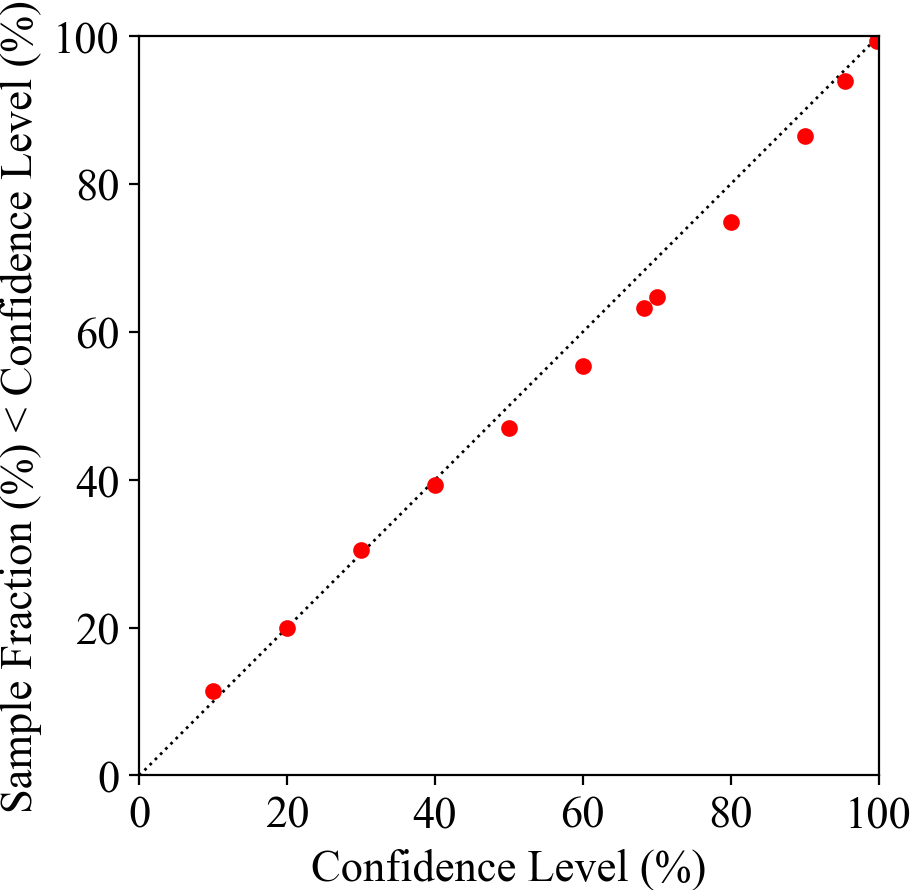}}
            \subfloat[][]{\includegraphics[width=4.0cm]{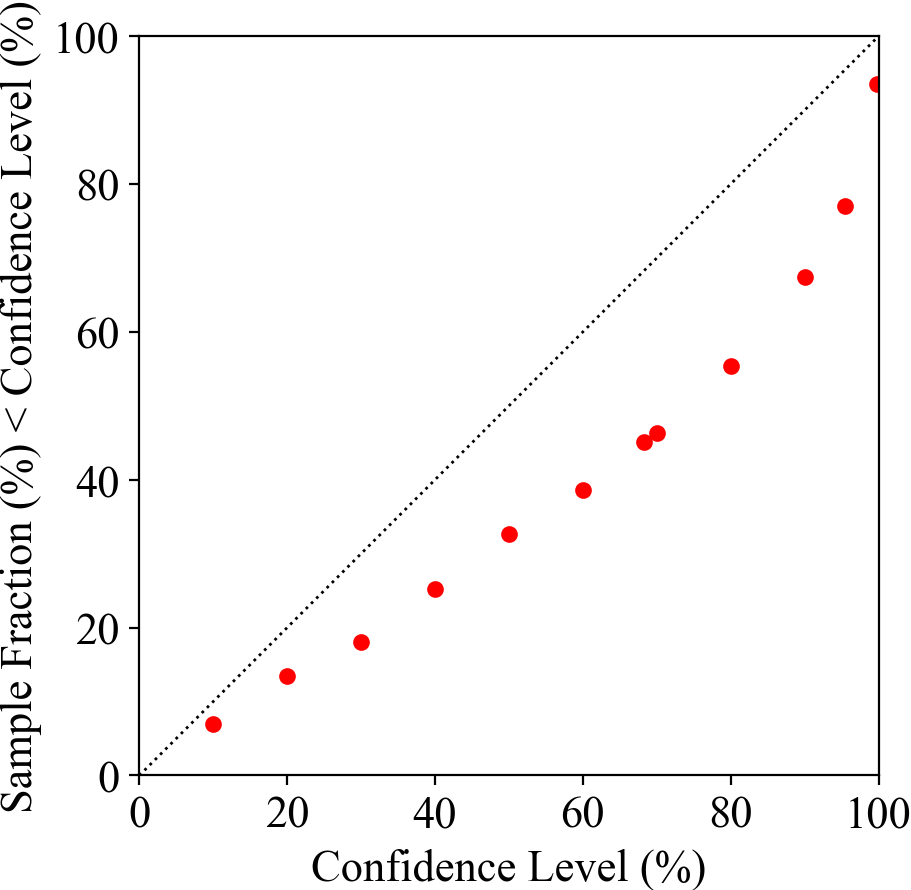}}
        \end{minipage}

        \begin{minipage}{0.24\linewidth}
            \centering
            \subfloat[][]{\includegraphics[width=4.0cm]{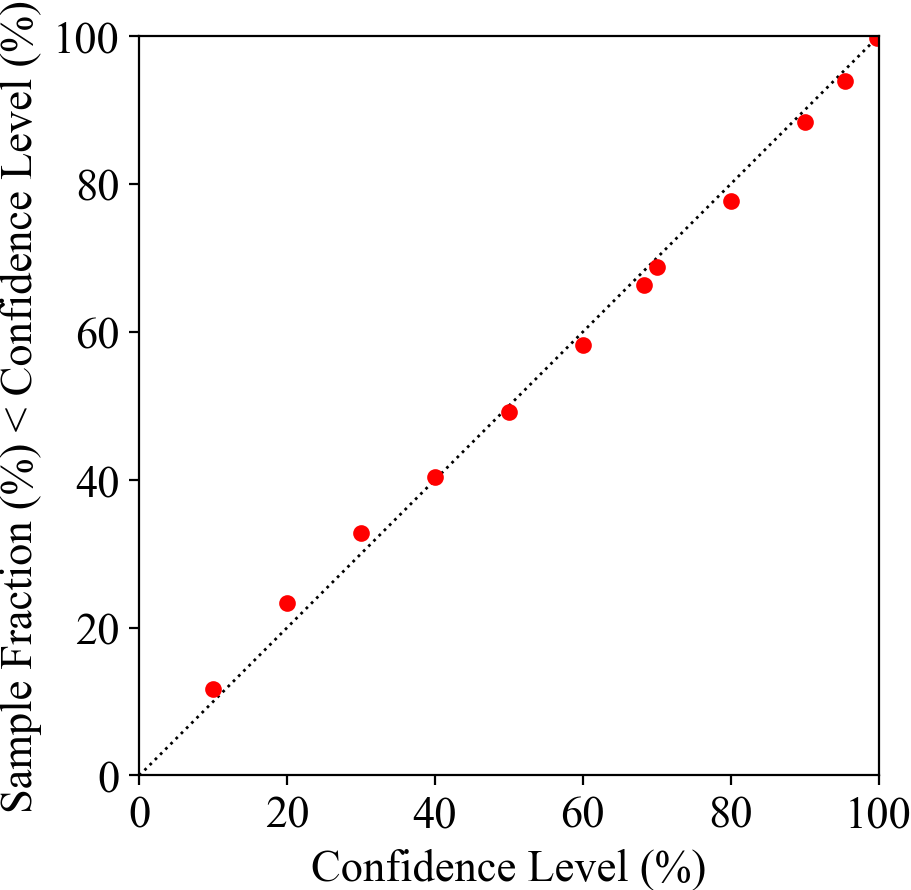}}
            \subfloat[][]{\includegraphics[width=4.0cm]{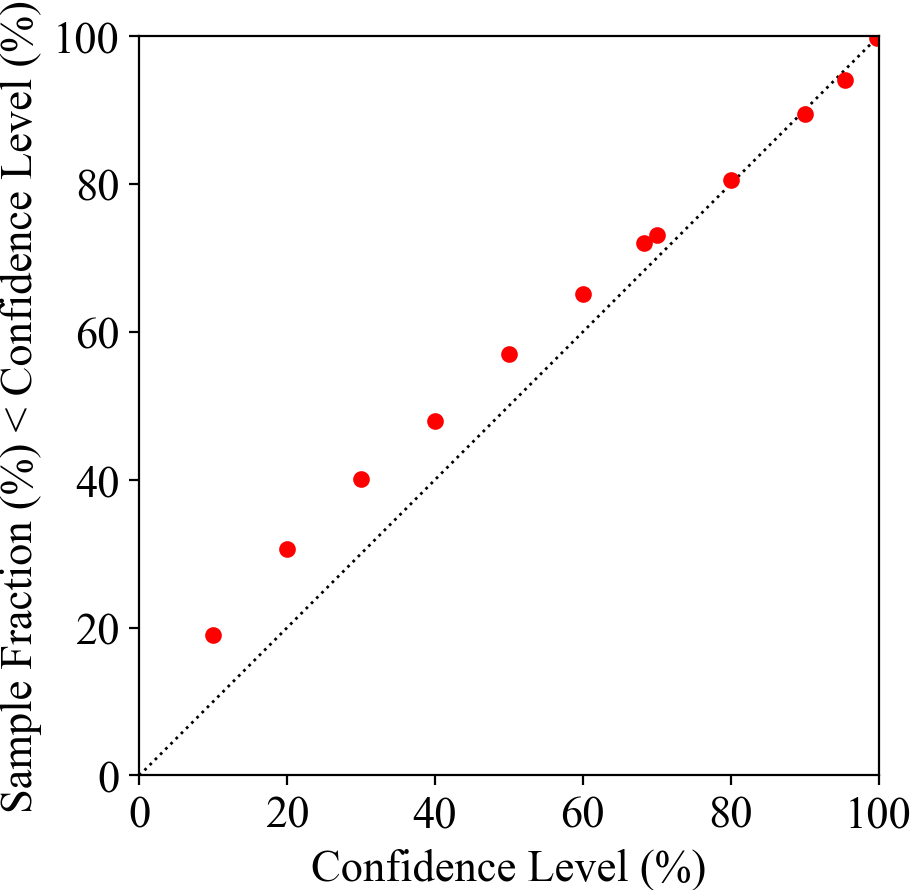}}
            \subfloat[][]{\includegraphics[width=4.0cm]{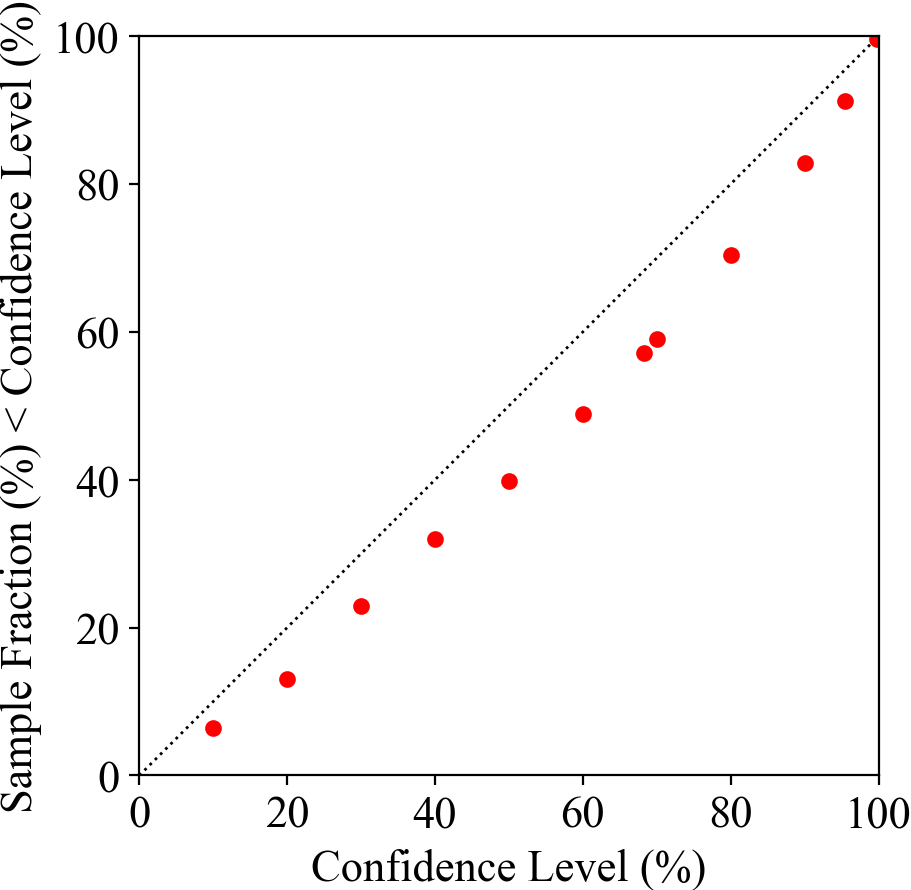}}
            \subfloat[][]{\includegraphics[width=4.0cm]{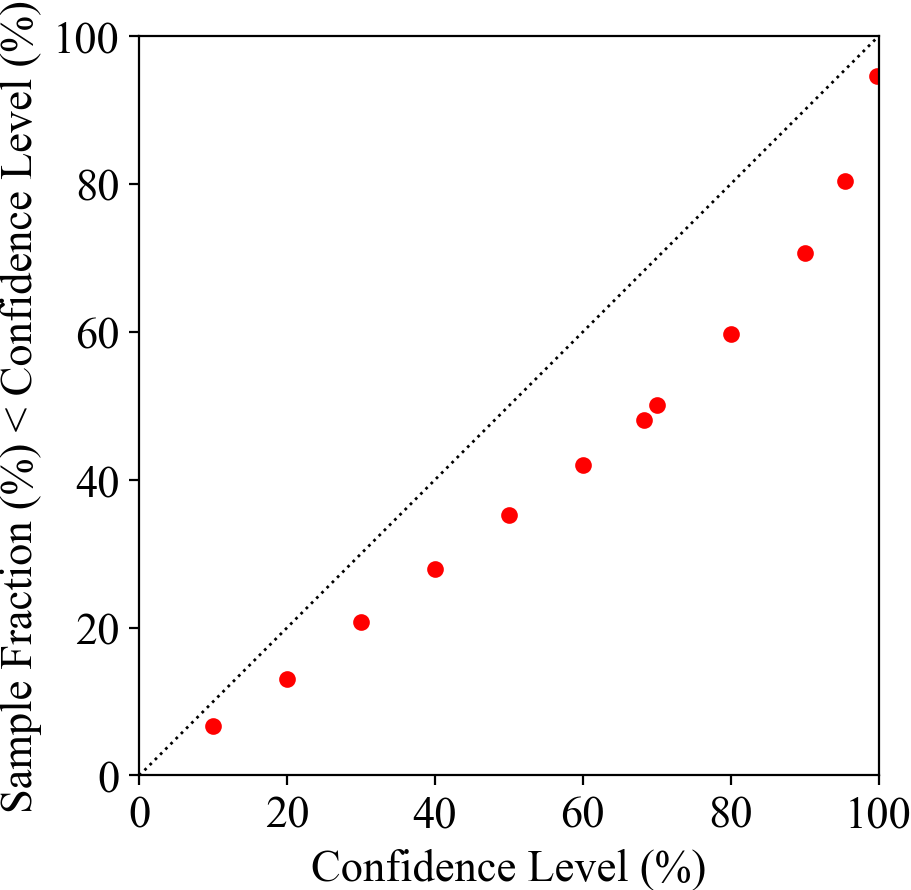}}
        \end{minipage}

        \caption{ Test results for source-dominating weak bursts. The upper panels show the statistical results obtained with different localization methods (see Table \ref{TABLE_Scheme}) at $\sim$ 570 total observed counts for 12 detectors (the brightest detector's observed counts is $\sim$ 80), including (a) $\mathcal B_{\rm POIS}$ (see Section \ref{Sec_Like_Pois}), (b) $\mathcal B_{\rm SG}$ (see Section \ref{Sec_Like_SG}), (c) $\chi_{\rm MIN}^{2}$ (see Section \ref{Sec_ChiSq_MIN}), (d) $\chi_{\rm GBM}^{2}$ (see Section \ref{Sec_ChiSq_GBM}). The medium panels show the statistical results obtained with the different methods mentioned in the upper panels at $\sim$ 100 total observed counts for 12 detectors (the brightest detector $\sim$ 20 counts) and the lower panels show those results at $\sim$ 20 total observed counts for 12 detectors (the brightest detector $\sim$ 3 counts). The localization statistical error is reasonable for all methods at $\sim$ 570 total observed counts, and all methods except for $\mathcal B_{\rm POIS}$ start to deviate one-one line at $\sim$ 100 total observed counts and then more severe at $\sim$ 20 total observed counts. $\mathcal B_{\rm SG}$ overestimates localization error regions while $\chi_{\rm MIN}^{2}$ and $\chi_{\rm GBM}^{2}$ underestimate localization error regions. However, $\mathcal B_{\rm SG}$ and $\chi_{\rm MIN}^{2}$ are more closer to one-one line than original DoL method $\chi_{\rm GBM}^{2}$. }
        \label{fig3}
    \end{figure*}

    For the source-dominating case (i.e. the background counts are less than the source counts), the background level of 0.5 counts/detector is adopted for simulations. As shown in Figure \ref{fig3}, when the total observed counts is $\sim 570$ for 12 detectors (the brightest detector's observed counts is $\sim$ 80), these four localization methods can obtain the correct localization map basically. But when the total observed counts decreased to $\sim$ 100 for 12 detectors (the brightest detector's observed counts is $\sim$ 20), all methods start to deviate from one-one line except for $\mathcal B_{\rm POIS}$. $\mathcal B_{\rm SG}$ and $\chi_{\rm MIN}^{2}$ are obviously closer to the one-one line than $\chi_{\rm GBM}^{2}$. When the total observed counts decreased to $\sim$ 20 for 12 detectors (the brightest detector's observed counts is $\sim$ 3), a correct localization map could also be obtained for $\mathcal B_{\rm POIS}$. We note that some very short duration ($\sim$ ms) bursts, e.g. the Terrestrial Gamma-ray Flashes \citep{TGF_BATSE_Fishman1994, TGF_GBM_Roberts2018}, could reach such low counts.

    From these tests, one can note that the performance of Bayesian methods is better than $\chi^{2}$ methods for the same inputs and settings. And the Bayesian method with Poisson likelihood is more applicable than Gaussian-based methods. Therefore, $\mathcal B_{\rm POIS}$ is recommended for localization. It should be noted that, as mentioned above, the simple Poisson likelihood (Equation \ref{EQ_Pois_L}) used in this paper should be replaced by the PGSTAT likelihood in real data for considering the background uncertainties \citep{Loc_BALROG_Michael2018}.

    For the original DoL method $\chi_{\rm GBM}^{2}$, its normalization factor $f_{i}$ is an approximate solution of $\chi^{2}$ minimization. As improvements, $\chi_{\rm MIN}^{2}$ conducts the minimization with a numerical solution, and $\mathcal B_{\rm SG}$ enhances it under the statistical framework of Bayesian. Thus in comparison with $\chi_{\rm GBM}^{2}$, the improvements of $\chi_{\rm MIN}^{2}$ and $\mathcal B_{\rm SG}$ could be seen in the tests as shown in the background-dominant and source-dominant weak bursts localization. Besides, due to the Gaussian assumption of $\chi^{2}$ localization methods, they could only be used for those cases with sufficient counts in detectors, which means the burst should not be too weak (say $<30\%$ of GRB 170817A) or too short (say $\sim$ ms). We note that the localization capabilities depend on the bursts' properties (e.g. spectrum), detector configuration, and incident angle, thus the above threshold for the correct location may vary with the instrument setting and incident angle of bursts.

    There are many settings and choices in the localization analysis that may potentially affect the final results. They are usually complex and coupled together, such as the selection of detectors, choice of spectral channels binning, spectral models, iteration termination criteria, etc. Some of them have been discussed in previous studies \citep[e.g.][]{Loc_BALROG_Michael2018, Loc_BALROG_Berlato2019}. Here we explore how to divide energy channels to optimize the localization results with simulations. To estimate the influence on the localization caused by the different data binning strategies in spectral channels, we did simulations for the medium bright burst with two kinds of binning data in the whole energy range (i.e. from 8 to 1000 keV): just one whole energy channel for each detector or divided to 8 energy channels for each detector. For the divided energy channels, the mathematical Poisson likelihood and logarithmic likelihood are:

        \begin{align}
            \mathcal L_{\rm P div}(i) = \prod \limits_{j} \prod \limits_{k} \frac{ \left( b_{j,k} + f_{i} \cdot m_{j,k,i} \right) ^ { s_{j,k} } \cdot \exp( -( b_{j,k} + f_{i} \cdot m_{j,k,i} ) ) }{ s_{j,k} ! }
            \label{EQ_Pois_P_div}
        \end{align}

        \begin{equation}
        \begin{aligned}
            \ln{ \mathcal L_{\rm P div}(i) } = & \sum_{j} \sum_{k} [ s_{j,k} \cdot \ln( b_{j,k} + f_{i} \cdot m_{j,k,i} ) \\
            & - ( b_{j,k} + f_{i} \cdot m_{j,k,i} ) -\ln s_{j,k}! ]
            \label{EQ_Pois_L_div}
        \end{aligned}
        \end{equation}
    where $s_{j,k}$ and $b_{j,k}$ are the total observed counts and background for energy channel $k$ in detector $j$, $f_{i} \cdot m_{j,k,i}$ is the expected source contribution in a single channel.

\begin{figure*}[ht]
        \flushleft

        \begin{minipage}{0.32\linewidth}
            \centering
            \subfloat[][]{\includegraphics[height=4.2cm]{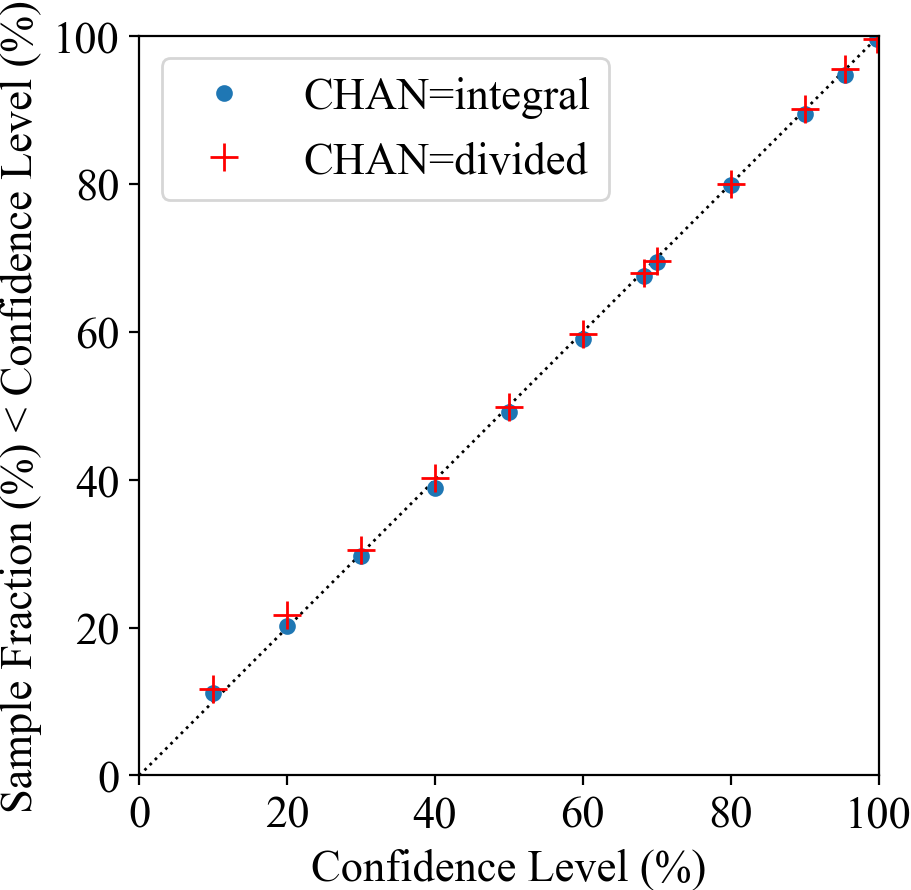}}
            \subfloat[][]{\includegraphics[height=4.2cm]{./Fig/pfig5_a1.png}}
            \subfloat[][]{\includegraphics[height=4.2cm]{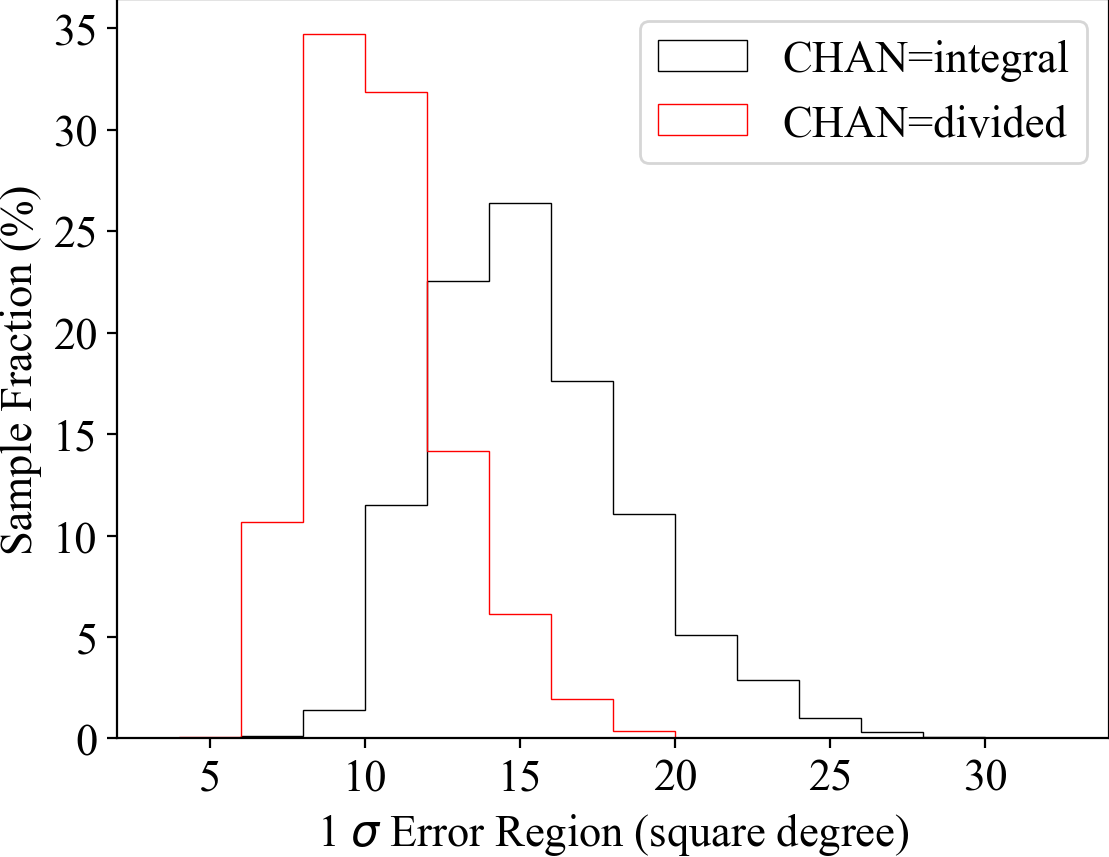}}
        \end{minipage}

        \begin{minipage}{0.32\linewidth}
            \centering
            \subfloat[][]{\includegraphics[height=4.2cm]{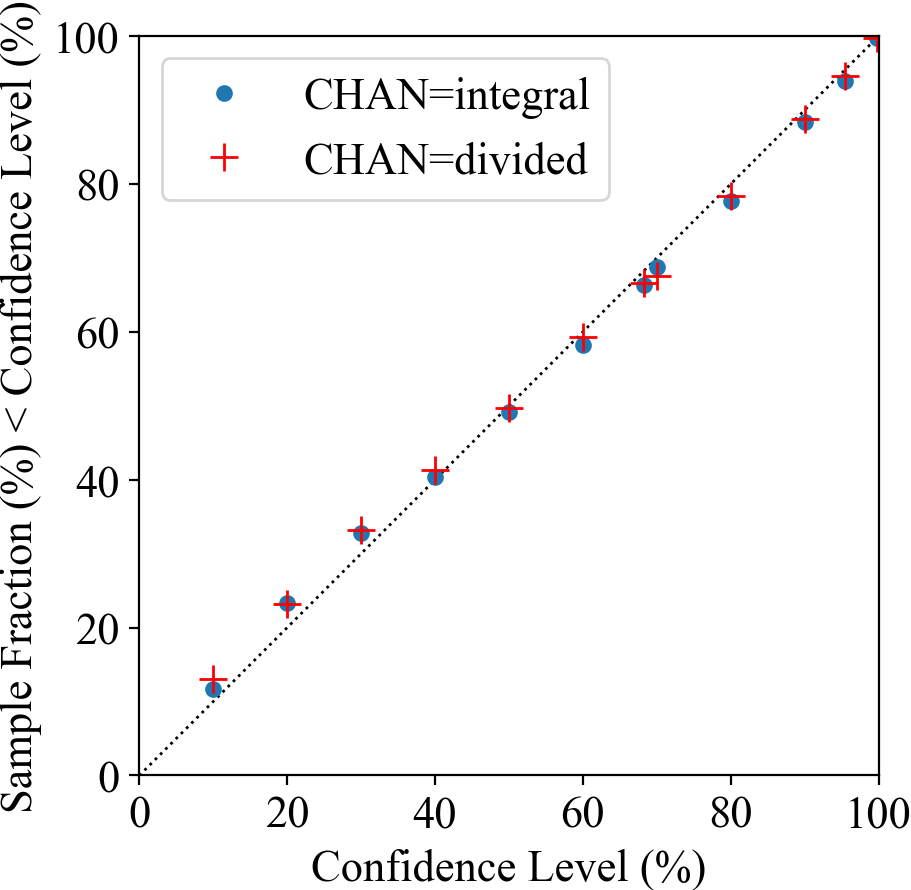}}
            \subfloat[][]{\includegraphics[height=4.2cm]{./Fig/pfig5_b1.png}}
            \subfloat[][]{\includegraphics[height=4.2cm]{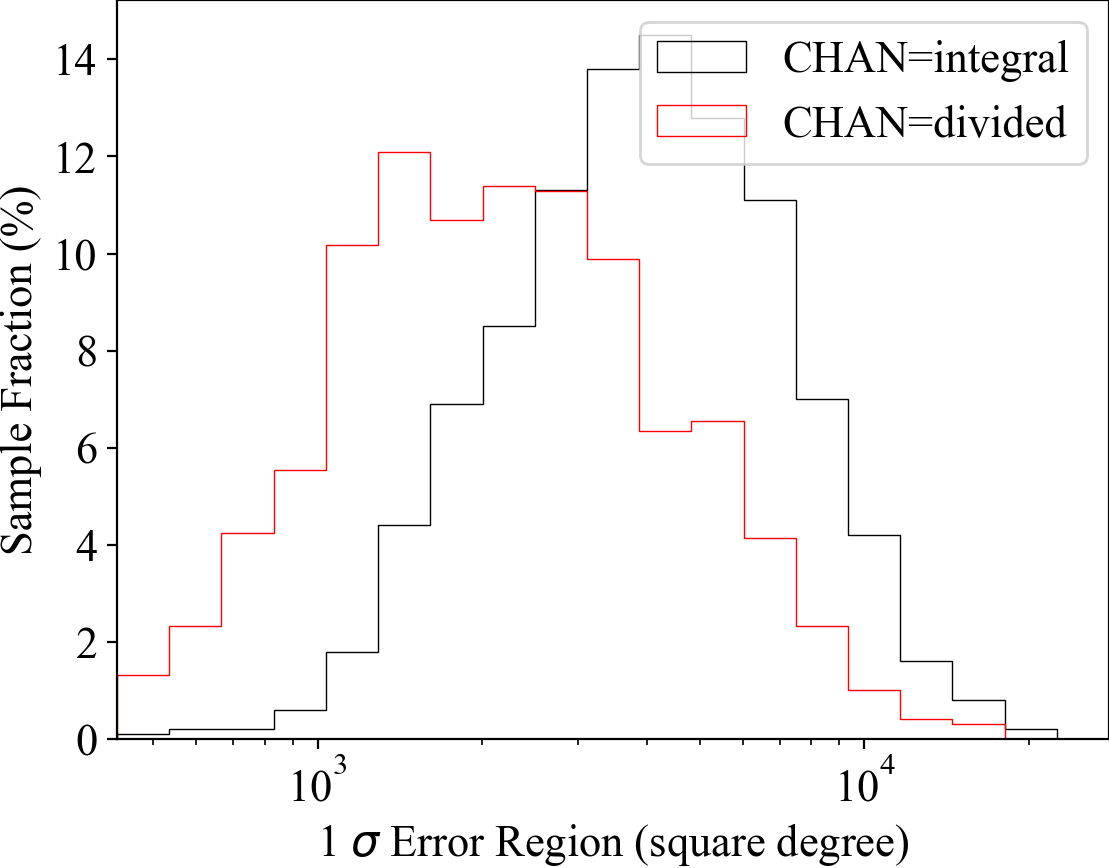}}
        \end{minipage}

        \caption{ Test results for localization setting of energy channels integral and divided with $\mathcal B_{\rm POIS}$ (see Section \ref{Sec_Like_Pois}). The upper panels show the comparisons of (a) the statistical results, (b) the distribution of offset between location center and truth location, and (c) the distribution of 68.27\% HPD error region for the medium bright source as shown in Table \ref{TABLE_SourceType}. The channels integral results are the same as Figure \ref{fig1} (a). The lower panels show the comparisons of (d) the statistical results, (e) the distribution of the offset between location center and truth location, and (f) the distribution of the 68.27\% HPD error region for the source-dominating weak bursts. The channels integral results are the same as Figure \ref{fig3} (i). The results indicate that the energy channels divided localization setting reduces the error region and offset between location center and truth location on the premise that the error obeys the statistics for the medium bright source and the source-dominating weak bursts. }
        \label{fig5}
    \end{figure*}

    As presented in Figure \ref{fig5} (a) to (c), the divided channels (i.e. 8 channels from 8 to 1000 keV) case can reduce the location center offset and error region with a factor of $\sim \frac{1}{3}$ compared to that of the integrated channel (i.e. 1 channel from 8 to 1000 keV) case. This tendency also could be found in the source-dominant weak bursts as shown in Figure \ref{fig5} (d) to (f).

\clearpage
\section{Location-Spectrum Iteration Localization}

    In real observation, using the fixed spectral templates is imperfect. An inevitable problem of the fixed templates localization strategy is that the spectra of preset templates usually differ from those of bursts, which may introduce substantial systematic errors.

\begin{figure*}[ht]
        \flushleft

        \begin{minipage}{0.45\linewidth}
            \centering
            \subfloat[][]{\includegraphics[height=5.5cm]{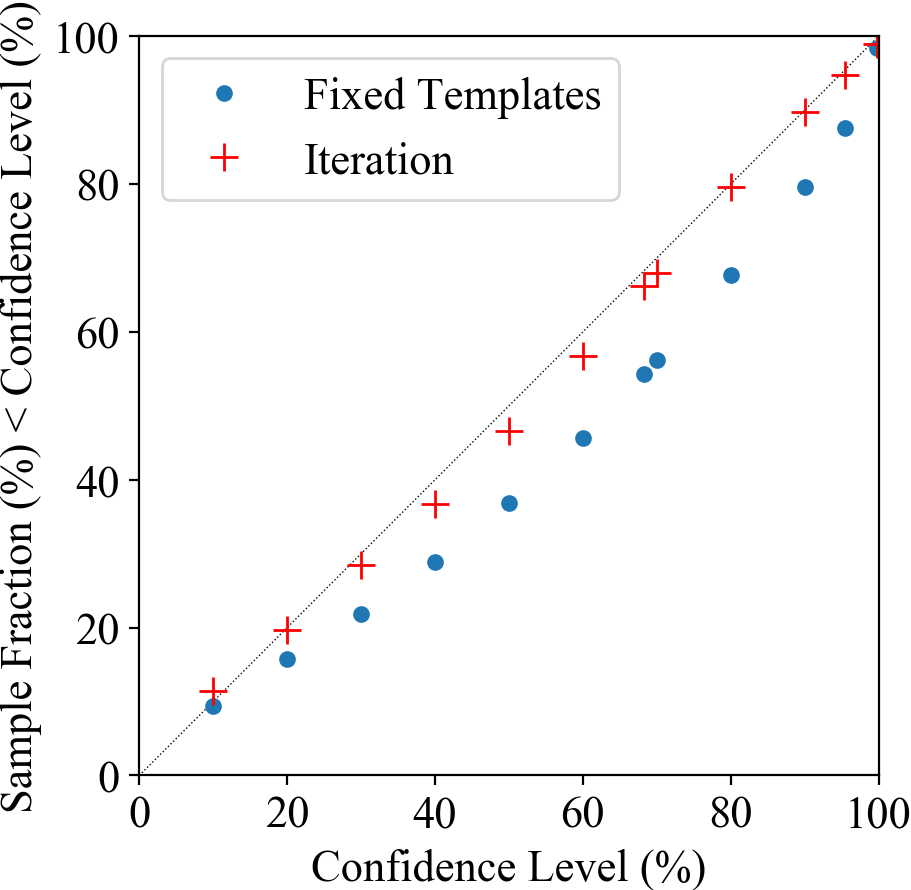}}
            \subfloat[][]{\includegraphics[height=5.5cm]{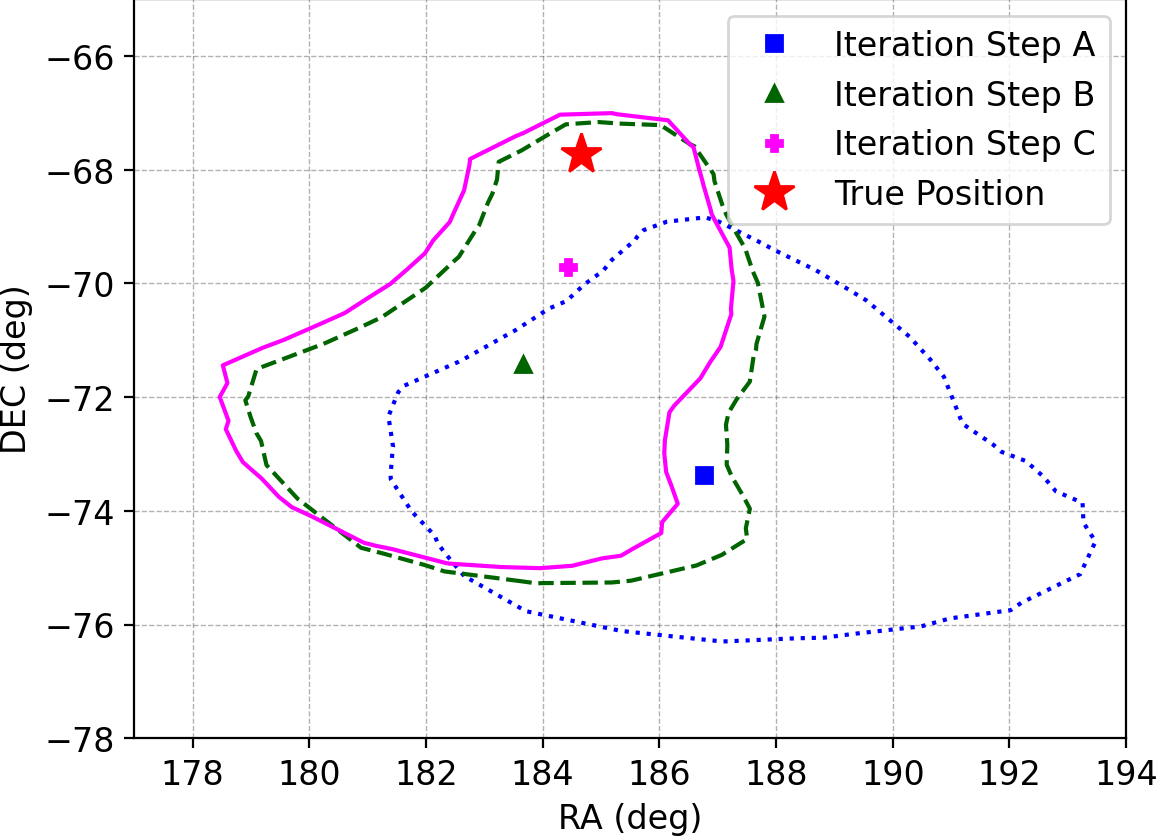}}
        \end{minipage}

        \caption{ Test results of the location-spectrum iteration localization. (a) Results for fixed templates method (blue) and location-spectrum iteration (red). (b) Evolution of the location results given by the location-spectrum iteration localization for a simulated medium bright burst (see also Table \ref{TABLE_Iteration}). The blue[green][magenta] represents the location center and the 68.27\% credible region of Step A[B][C], respectively. Other captions are the same as Figure \ref{fig1}. }
        \label{fig2}
    \end{figure*}

    To illustrate the performance of the fixed templates localization strategy, i.e. the deviation of location induced by the difference of spectrum, a simulation using $\mathcal B_{\rm POIS}$ localization method with fixed templates for the medium bright burst is implemented. As shown in Figure \ref{fig2}, the statistical distribution of the location results significantly deviates from the expected value with the error region being underestimated.

    Since the fixed template localization has advantages in the calculation, to alleviate the above issues of fixed templates localization, we propose a location-spectrum iteration approach for $\mathcal B_{\rm POIS}$ localization method, as described below:

    \begin{itemize}

        \item[\textbf{Step A:}] First, derive the initial localization result with preset localization templates, which is similar to previous studies \citep{Loc_GBM_Connaughton2015, Loc_GBM_Goldstein2020}, but implemented by $\mathcal B_{\rm POIS}$ (i.e. the Bayesian localization method with fixed spectral template).

        \item[\textbf{Step B:}] With the initial location from Step A, spectral analysis of the burst is implemented with all detectors. Then redo the localization with the updated location template which is calculated based on the burst spectrum. Subsequently, iterate the spectral fitting and localization. The purpose of selecting all detectors for the spectral fitting here is to account for a large deviation of the location. To achieve an appropriate spectrum, four spectral models are employed to fit independently: the Band function, the Comptonized, power law, and power law + black body model. The best model is selected by the Bayesian information criterion (BIC) \citep{common_STAT_BIC9}. This iteration will terminate if the observed counts are consistent with the expected counts in 90\% of the C-statistics confidence level or if iterating more than several times. Here four times are selected for simulated tests.

        \item[\textbf{Step C:}] Based on the location obtained from Step B, a refined spectral analysis is executed with a sample of good detectors which are selected based on preset criteria, including incident angle \textless $60^{\circ}$ and significance \textgreater $5~\sigma$. With the refined spectrum, the template spectrum will be updated and the final localization result is obtained. The terminated condition is the same as in Step B.

    \end{itemize}

    To quantitatively estimate the performance of this location-spectrum iteration localization strategy, an MC simulation has been implemented. As shown in Figure \ref{fig2}, the localization result of the iteration strategy is significantly improved (compared to that of the fixed templates) and thus quite consistent with the expectation. Taking a burst for example, we traced the evolution of the spectral and location parameters during the iteration, as shown in Table \ref{TABLE_Iteration} and Figure \ref{fig2} (b). Both the spectrum and location center tend to converge to the true value as iteration goes. Although the final spectrum (derived in Step C) is not exactly the same as the input one (which is not surprising if think of the spectral fitting error), the final location map given by this method is statistically reliable and correct according to the statistical validation in the left panel of Figure \ref{fig2}.

    %

    \begin{table*}[ht]
        \centering
        \caption{ Parameter evolution of the location-spectrum iteration for a simulated medium bright burst (Table \ref{TABLE_SourceType}), which is also shown in Figure \ref{fig2} (b). }
        \label{TABLE_Iteration}
        \begin{tabular}{ccccc}
            \hline
            Steps & Best Spectral Model & Index & $E_{\rm peak}$ (keV) & Offset (deg) \\
            \hline
            Input  & Comptonized &  -1.50         &  200       & -    \\
            Step A & Comptonized &  -1.15         &  350       & 5.68 \\
            Step B & Comptonized & $-1.15\pm0.27$ & $230\pm38$ & 3.71 \\
            Step C & Comptonized & $-1.33\pm0.13$ & $187\pm21$ & 2.00 \\
            \hline
        \end{tabular}
    \end{table*}

\clearpage
\section{Discussions and Conclusions}

    In this paper, we propose a Bayesian localization method that can not only give more accurate results but also can be used for scenarios when the computational resource is constrained or the calculation speed is preferred, such as the in-flight localization or low-latency localization for rapid follow-up observations. 

    Take the in-flight localization for example, the computing resources and memory on board are very limited, and the in-flight localization software (e.g. \textit{Fermi}/GBM, GECAM) usually employs $\chi^{2}$ minimization method with a coarser HEALPix pixel (i.e. 5.0$^{\circ}$ apart which means 8112 HEALPix pixels number) and only one localization template. If using our Bayesian method as in-flight localization, it only takes $\sim$ 20 s to give location results, which is much acceptable compared to the MCMC-based localization algorithm developed for GBM and GECAM which will take typical $\sim$ 20 minutes \citep{Loc_MCMC_Liao2020}.

    Comparison between our method and the $\chi^{2}$ minimization method are studied in detail. This comparison was done with dedicated simulations which eliminate bias and impacts introduced by the inaccuracies in detector response, background estimation, and knowledge of burst spectrum. The reliability and correctness of the location results are validated by directly checking the confidence regions of the localization probability map through comprehensive simulations.

    We find that, for medium-bright bursts, all four kinds of localization methods studied in this paper give similar and reliable location results. But for source-dominant weak bursts, only $\mathcal B_{\rm POIS}$ could give a correct localization probability map which is useful for some short-duration bursts, i.e. TGFs. For background-dominant weak bursts, $\mathcal B_{\rm POIS}$ and $\mathcal B_{\rm SG}$ could give a correct localization probability map at \textgreater 90\% HPD credible region. Therefore, the Bayesian method with Poisson likelihood is recommended rather than the $\chi^{2}$ minimization method. In real observations, the more sophisticated PGSTAT statistic should be used instead of the simple Poisson likelihood \citep{YiZhao_LOC_GECAM}.

    We also find that compared to the original DoL method $\chi_{\rm GBM}^{2}$, $\chi_{\rm MIN}^{2}$ and $\mathcal B_{\rm SG}$ could improve the localization results by the numerical solution during maximization and utilization of the Bayesian inference. We demonstrate that the mismatch of the burst spectrum and template spectrum will cause location deviation, which may increase the systematic error of localization.

    We also proposed a Bayesian-based location-spectrum iteration localization method to take advantage of and alleviate the issues of the fixed spectral template method. Compared to the existing methods for optimizing spectrum (i.e. fixed templates and location-spectrum simultaneously fitting), our location-spectrum iteration localization strategy features the following advantages: (1) the mismatch between the spectrum of fixed localization templates and burst spectrum could be fairly eliminated, and (2) the calculation of localization process is straightforward. Thus the location-spectrum iteration localization method which requires few computing resources has the potential to deploy onboard as the in-flight localization software. Indeed, the one-time location-spectrum iteration localization has been used for GECAM (Huang et al. 2023, in press; Zhao et al. 2023, in press). However, it may still have convergence problems during iteration in some cases. Besides, we find that dividing the counts into several bins would improve the localization results than treating these counts as a single bin. However, these localization settings usually require more memory and computing resource which means that they may be not suitable for in-flight localization software.

    At last, we note that there are some open questions in the localization, e.g., how to get a reliable localization probability map for background-dominant weak bursts. As mentioned in Section 3, the current methods studied in this paper are not able to give reliable location error regions $<$ 90\% confidence level for weak bursts. On the other hand, these weak bursts might be very important as they could be associated with Gravitational Waves or Fast Radio Bursts, thus a joint time-delay localization with multiple all-sky instruments is highly required to provide reliable location results.

\normalem
\begin{acknowledgements}
    This work is supported by the National Key R\&D Program of China (2021YFA0718500). We thank the support from the Strategic Priority Research Program on Space Science, the Chinese Academy of Sciences (Grant No. XDA15360102, XDA15360300, XDA15052700), the National Natural Science Foundation of China (Grant No. 12173038, U2038106) and the National HEP Data Center (Grant No. E029S2S1). S.L.X. appreciate the \textit{Fermi}/GBM team for public data, opened software, and helpful discussions. We thank Li Chen (Beijing Normal University), Xi Long (Harvard University), Merlin Reynaard Kole (University of Geneva), and Nicolas Produit (University of Geneva) for helpful discussions.
\end{acknowledgements}


\end{document}